\documentclass[aps,reprint,pre]{revtex4-1}

\usepackage{graphicx}
\usepackage{amsmath}
\usepackage{amssymb}
\usepackage{commath}
\usepackage{siunitx}
\usepackage{xcolor}

\newcommand{\beginsupplement}{%
        \setcounter{table}{0}
        \renewcommand{\thetable}{S\arabic{table}}%
        \setcounter{figure}{0}
        \renewcommand{\thefigure}{S\arabic{figure}}%
     }

\begin{document}


\title{Self-organisation of tip functionalised elongated colloidal particles}



\author{Mariana Oshima Menegon}
\email[Corresponding author: ]{m.oshima.menegon@tue.nl}
\affiliation{Department of Applied Physics, Eindhoven University of Technology, PO Box 513, 5600\,MB Eindhoven, The Netherlands.}

\author{Guido L. A. Kusters}
\affiliation{Department of Applied Physics, Eindhoven University of Technology, PO Box 513, 5600\,MB Eindhoven, The Netherlands.}

\author{Paul van der Schoot}
\affiliation{Department of Applied Physics, Eindhoven University of Technology, PO Box 513, 5600\,MB Eindhoven, The Netherlands.}
\affiliation{Institute for Theoretical Physics, Utrecht University, Princetonplein 5, 3584\,CC Utrecht, The Netherlands.}


\date{\today}

\begin{abstract}
Weakly attractive interactions between the tips of rod-like colloidal particles affect their liquid-crystal phase behaviour due to a subtle interplay between enthalpy and entropy.
Here, we employ molecular dynamics simulations on semi-flexible, repulsive bead-spring chains of which one of the two end beads attract each other. We calculate the phase diagram as a function of both the volume fraction of the chains and the strength of the attractive potential. We identify a large number of phases that include isotropic, nematic, smectic A, smectic B and crystalline states. 
For tip attraction energies lower than the thermal energy, our results are qualitatively consistent with experimental findings: we find that an increase of the attraction strength shifts the nematic to smectic A phase transition to lower volume fractions, with only minor effect on the stability of the other phases. 
For sufficiently strong tip attraction, the nematic phase disappears completely, in addition leading to the destabilisation of the isotropic phase. 
In order to better understand the underlying physics of these phenomena, we also investigate the clustering of the particles at their attractive tips and the effective molecular field experienced by the particles in the smectic A phase. 
Based on these results, we argue that the clustering of the tips only affects the phase stability if lamellar structures (``micelles'') are formed. We find that an increase of the attraction strength increases the degree of order in the layered phases.
Interestingly, we also find evidence for the existence of an anti-ferroelectric smectic A phase transition induced by the interaction between the tips. A simple Maier-Saupe-McMillan model confirms our findings. 
\end{abstract}


\maketitle 
\section{Introduction}
\label{sec:intro}
Elongated colloidal particles form additional phases under conditions in between those where the well-known isotropic (disordered) and crystalline (ordered) phases are found \cite{Kuijk2012, Meyer1990, Furukawa1993}. 
The particles are invariably aligned but have no or only partial (short-ranged or quasi long-ranged) positional order in these phases and, for this reason, are called liquid-crystalline phases. 
The phase transitions are driven primarily by entropy, as theoretical, simulation, and experimental studies have shown \cite{Odijk1985,Odijk1986,Chen1993,Hidalgo2005,Shundyak2006,Grelet2008hex,Grelet2014,Grelet2016}. 
More recently, the use of selective surface functionalisation of elongated colloidal particles has opened up an interesting novel path of investigation, allowing us to modify the self-assembled liquid-crystalline phases and/or to manipulate their stabilities. 
For example, such particles have been explored in the synthesis of functional materials including nanowires and batteries \cite{Nam885,Tseng2006,Steinmetz,Lee1051}, and in the investigation of specific structure formation such as multipods, tubes, and bottle brushes \cite{Chaudhary2012,Cotte2017,Park348,Gao2018}. 
Nevertheless, studies concerning how a relatively weak and highly local surface modification affects the phase sequences for a wide range of concentrations, as well as particles characteristics such as aspect ratio and bending flexibility, remain scarce. 
 
A good example of functionalised elongated colloids is the recent work by Repula \emph{et al.} using filamentous M13 virus particles \cite{PatchyRef}, which measure 1\,\si{\micro\meter} in length, 7\,nm in width and which have persistence length of about 3\,\si{\micro\meter}. 
In their experiments, the M13 virus has its terminal (P3) protein modified, allowing for the attachment of red dye molecules to one of the tips of these polar particles. 
The procedure results in a controllable, attractive, single-end local interaction in what was previously a purely repulsive rod. From previous work, we know that the complete phase sequence of suspensions of such viruses comprises isotropic, nematic, smectic A, smectic B, columnar, and crystalline phases \cite{Grelet2014}. 
Interestingly, the surface modification seems to affect only one of various phase transitions: the nematic-smectic A phase transition is influenced by the number of red dye molecules grafted to the virus tip, stabilisesing the latter phase. 
For the purpose of understanding the reason for this, we investigate how a weakly attractive tip modifies the liquid crystalline behaviour of repulsive, semi-flexible rod-like particles using computer simulations.

In this paper, we present the calculated phase diagram of such particles as a function of both the concentration and the attraction strength between tips, demarcating two regimes. 
In the first regime, in which the strength of the tip-tip interaction corresponds to energies lower than or comparable to the thermal energy, our results are qualitatively consistent with the experimental findings \cite{PatchyRef}. 
In the second regime, corresponding to slightly stronger attractive energies, we find interesting effects including the complete suppression of the nematic phase and the destabilisation of the isotropic phase. 
Additionally, we address in this paper several other topics regarding the microstructure of the phases, such as: (1) qualitative and quantitative aspects of the supramolecular aggregation due to the presence of the attractive tips in the various phases; (2)  the response of the interlayer distance and the molecular field for the various concentrations and attraction strengths between the tips in the smectic and crystalline phases; (3) evidence for the existence of an anti-ferroelectric smectic A phase induced by the interaction between the ends. 
We note in this context that the existence of an anti-ferroelectric phase of end-functionalised hard rods with double periodicity was anticipated long ago by Jackson and co-workers by means of density function theoretical calculations \cite{Sear1995}. We also present a simple model based on the Maier-Saupe-McMillan theory that describes the phase transition within the smectic A phase and that qualitatively explains our simulation results.

The remainder of this paper is structured as follows.
We describe the methods and model particle employed in our computer simulations as well as the data analysis procedure in Section \ref{sec:met}. In Sections \ref{sec:phase} and \ref{sec:mic}, we present the results followed by a discussion. These sections are devoted to the phase behaviour and the microstructure of the phases. In Section \ref{sec:con}, we present our most important conclusions. Finally, in the Appendix, we present our Maier-Saupe-McMillan theory for end-functionalised, perfectly parallel rods.

\section{Methods and analysis}
\label{sec:met}

\emph{Model particles - }We model the semi-flexible rod-like particles as bead-spring chains of $n$ overlapping beads of diameter $D$ and mass $m$. 
Consecutive pairs of beads interact via a harmonic potential $U_{r}=\kappa_r(r-D/2)^2/2$, which means that the beads overlap by a half diameter. 
Here, $r$ is the bond length and $\kappa_r$ is the harmonic bond stretch constant that we fix to a reasonably high value of $100\,k_\mathrm{B}T/D^2$, where $k_\mathrm{B}T$ denotes the thermal energy with $k_\mathrm{B}$ Boltzmann's constant and $T$ the absolute temperature. 
Consecutive bonds between beads interact via a harmonic bending potential, $U_{\theta}=\kappa_\theta(\theta-\pi)^2/2$, where $\theta$ is the angle between two consecutive bonds and $\kappa_\theta$ the harmonic bend constant.
Except for the first and second neighbours within a chain, the interactions between the beads are given by a steeply repulsive potential for which we use the Lennard-Jones potential, truncated at its minimum and shifted to zero, i.e., $U_R=4\epsilon_0[(R/D)^{-12}-(R/D)^{-6}]+\epsilon_0$ if $R\leqslant2^{1/6}D$ and $U_R=0$ if $R>2^{1/6}D$, where $R$ is the centre-to-centre distance between the beads and $\epsilon_0$ the strength of the interaction, which is kept constant and equal to $1\,k_\mathrm{B}T$.
The single-end attractive interaction is modelled using a second type of bead at one of the ends of every chain, represented in red in the snapshots of our simulations (Figure \ref{imgSnap}). These beads interact with each other via the full Lennard-Jones potential, $U_\mathrm{LJ}=4\epsilon[(R/D)^{-12}-(R/D)^{-6}]$, with various values of $\epsilon\geqslant{0}$. Notice that because only one end of every chain is sticky, the chains are polar and lack inversion symmetry. 

\emph{Particle characteristics - }Due to the harmonic bond stretch, the contour length of the particles, $L$, and, consequently, their mean aspect ratio, $L/D=(n-1)r/2$, are slightly variable. 
We refer to the mean aspect ratio at very low densities, $L_0/D$, to characterise our particles. The actual contour length of the particles is somewhat smaller than this, in particular in the more congested phases due to the high ambient pressure that compresses the particles somewhat. See Reference \cite{Braaf2017} for a discussion.
The persistence length $L_P$ of the particles depends on the harmonic bend constant. We have $L_P=\kappa_\theta{r}/k_\mathrm{B}T$, at least for an infinitely large number of beads and $\kappa_\theta{r^2}/k_\mathrm{B}T\gg1$ \cite{Naderi2014}. 
The value for the end-to-end distance of particles of various persistence lengths obtained in our previous simulations compares well with the value of the end-to-end distance predicted by the worm-like chain model in the isotropic phase, showing that the relation is indeed appropriate to describe the persistence length of our particles even though they are not infinitely long \cite{Braaf2017}. 
We quantify the particles' flexibility by the ratio of the dilute-solution contour length and the persistence length, $L_0/L_P$. 
Simulations are performed for chains of aspect ratio $L_0/D=10.77$ and flexibility $L_0/L_P=0.3$.
The aspect ratio chosen gives us a reasonable compromise between equilibration time and particle number in simulations at high concentrations. 
The flexibility matches the one of the experimental model particles mentioned earlier \cite{PatchyRef}. More details are given in Section \ref{sec:phase}. 

In the supplementary material \footnote{See Supplemental Material at the end of this file for preliminary results for shorter and stiffer chains of $L_0/D=6.46$ and $L_0/L_P=0.1$ as well as the initial configurations employed in the expansion simulations and, again for the longer and more flexible chains of $L_0/D=10.77$ and $L_0/L_P=0.3$, further analysis of the average aggregation number and a comparison between the smectic order parameter for two values of the attraction strength. All files related to a published paper are stored as a single deposit and assigned a Supplemental Material URL.}, we in addition present preliminary results for somewhat shorter and stiffer chains of $L_0/D=6.46$ and $L_0/L_P=0.1$ . 
The only impact aspect ratio seems to have is that the volume fractions at which the various phase transitions take place decreases as the aspect ratio increases, as in fact expected from previous studies \cite{Braaf2017}. The same is true for the effect of bending flexibility, which increases the volume fraction at which the various phase transitions occur.
Hence, we focus our presentation in this work on results for the longer particles.

\emph{Molecular dynamics (MD) - }We perform MD computer simulations on 4608 bead-spring chains in a box in order to obtain structural properties of the tip attractive rod-like particles.
Our simulations run for 20000 time units that, in physical quantities, correspond to $\sqrt{mD^2/\epsilon_0}$. We employ time steps of $10^{-3}$ in these units, in other words, a total of $2\times10^{7}$ time steps. We save configurations every $2\times10^{5}$ time steps. The volume $V$ of the box is equilibrated for fixed temperature $T$ and pressure $P$, using the Nos\'e-Hoover thermo- and barostat, which allows the box dimensions to adjust independently \cite{Plimpton1995}.
This choice is important to accommodate layered phases and obtain their undistorted interlayer distance. From the average equilibrated volume $V$, we calculate the volume fraction $\phi$ with the expression $\phi=Nv_0/V$, where $v_0$ is the volume of the spherocylinder whose length is equivalent to the average length of the chains at very low densities, $v_0=\pi{D^3}/6+\pi{D^2}L_0/4$. 
That means that the actual volume fraction is slightly lower than the value we adopt to represent it. 
This choice is made in order to keep most of the parameters of our simulation fixed.
The results we present in this paper are from expansion simulation runs starting from a crystal-like configuration, which we describe in more detail in the paragraph below. 
We also perform compression runs for the particles without attractive end tips for all phase transitions identified. These results are not shown. In these runs, the initial configuration is obtained from an (expansion) equilibrium configuration at a lower pressure $P$ (or volume fraction $\phi$) near the phase transition. From the compression runs, we note that the phase transitions take place at the pressure we expected from the expansion simulations. The largest disagreement in the calculated volume fraction of the resulting more condensed phase is 0.5\%. For this reason, our simulations starting from the crystal-like configuration seem to be robust.

\emph{Initial configurations - }For the expansion simulation runs, we consider diverse variations of the crystal-like initial configurations corresponding to how our polar particles are oriented. In all configurations, the particles are organised in 16 AAA stacked layers. In each layer, particles are organised in a hexagonal lattice and aligned parallel to $z$-direction. For this reason, if particles in our simulations have a preferential direction, which is described by the director $\mathbf{n}$, this is usually approximately parallel to the $z$-direction, $\mathbf{n}\parallel\mathbf{\hat{z}}$.
In the first type of initial configuration, all attractive beads are in the upper tip of the particle. 
Even though our simulation time is relatively long, it is not sufficiently long to equilibrate the system at all concentrations. 
In other words, in the more condensed phases, we do not reach 1:1 chains up and down relative to the director. 
In the second type of initial configuration investigated, the attractive beads alternate between the upper and the bottom tip within the same layer. 
In the third type of initial configuration tested, half of the layers have the attractive groups in the upper tip and the other half have the particles' attractive bead in the bottom tip. These layers alternate in a way that this third configuration has a bilayer type of structure. 
The latter two initial configurations result in structures with fewer or no defects at all. Hence, in the following sections, we present the results from the simulations in which the third type of initial configuration is used.  

\emph{Phase classification - }Our equilibrium configurations are classified using order parameters (OPs) and correlation functions. The liquid-crystalline phases we first distinguish from the isotropic phase by quantifying the degree of alignment of the particles through the usual nematic OP. A second OP quantifies the organisation of the particles in layers. If layers are not formed, we have either a nematic or a columnar phase. If layers are formed, we have either smectic or crystalline phases. A third OP quantifies the hexagonal bond order. This procedure allows us to distinguish between nematic and columnar, and between the smectic A and the smectic B and crystal phases. This final classification is possible by studying the correlation of the bond order (measuring two-dimensional hexagonal order) and the pair correlation function (probing radial in-plane positional order). Using this procedure, we are able to distinguish between all known phases of the fd virus suspensions, mentioned in Section \ref{sec:intro}. Our analysis procedure is similar to the one employed in our previous work to which we refer for details \cite{Braaf2017}. We also identify two distinct smectic and crystalline phases due to the attraction between the functionalised tips. The procedure to distinguish between them we describe below in the paragraph dealing with the anti-ferroelectric phase transition.

\emph{Aggregation statistics - }Further analysis is required in order to study the structure of supramolecular assemblies in all phases, in particular in the layered phases where the spacing is expected to be influenced by (i) the attraction strength between tip beads and (ii) the concentration of the particles. 
We quantify the aggregation of the particles by focusing on cluster sizes. 
By \emph{cluster} we refer to groups of attractive tip beads that are spatially close or connected by a bead satisfying this criterion. We arbitrarily choose a radius $r_c$. The first choice for $r_c$ is $1.24D$. This value corresponds to the second root of a parabola, whose minimum is at the minimum of the Lennard-Jones potential and the first root coincides with the root of this potential.
If pairs of attractive tip bead are closer than this distance, we consider them as belonging to the same group. Note that not all beads in a group need to be closer than $r_c$, as some can be connected indirectly via other attractive beads, as represented in the inset of Figure \ref{imgCluster10}, in Section \ref{sec:mic}.

\emph{Anti-ferroelectric phases - }Due to the polarity of our molecules, we identify two types of organisation in layered phases: one in which the attractive tip beads are present in roughly equal numbers every interlayer spacing and another in which the attractive tip beads are mostly present in every other interlayer spacing, creating a bilayer structure with double periodicity of that of the layers. 
In order to pinpoint the transition between these two states within our phase diagram, we project the particle orientations on a vector along the director. We define the orientation vector of the particles connecting the repulsive end bead to the attractive end bead. The two extreme situations are: (1) the rods are in a bi-layer type of configuration and, as a result, all particles within a layer are either parallel or anti-parallel to a vector along the director; (2) the ratio between parallel and anti-parallel particles to the same vector within a layer is 1:1. For these cases, if the fractions of particles in a particular orientation in even and odd layers are respectively $f=0$ and 0.5, then the anti-ferroelectric order parameter is defined as $\delta=1-2f$. We choose the value $\delta=0.5$ as a criterion to classify the phases as the usual smectic A ($\delta<0.5$) or the anti-ferroelectric smectic A$_2$ ($\delta\geqslant{0.5}$). The same procedure is applicable to the smectic B and crystalline phases, whose corresponding anti-ferroelectric phases are denoted smectic B$_2$ and crystalline$_2$ phases. 
From snapshots of our simulations, we find that even at the lowest attraction strength investigated, the smectic B and crystalline phases are anti-ferroelectric.  

\emph{Layer thickness - }We utilise two different procedures to calculate the interlayer distance in the layered phases. As a first procedure, 
we take the value that maximises the Fourier component of the normalised distribution of centre of masses along the director \cite{Polson1997}.
This does not differentiate between the layers regarding the fraction of particles pointing up. Because there are two types of smectic A phase, we expect that the interlayer distance is different if in their interface there are mostly attractive or repulsive tips. In order to measure this difference in the bi-layered phase, we plot the histogram of counts of the centre of mass for each (arbitrary) interval of positions along the director. 
In the layered phases, the distribution of the centres of mass of the particles along the director is peaked around the centres of the layers. 
Hence, we fit a Gaussian function to each counting of the centre of mass divided by the maximum count,
$g_j(z)=e^{-b_j(z-a_j)^2}$, 
where $e$ is the usual Euler constant. The parameter $b_j$ is related to the standard deviation $s$ of the Gaussian function by $b_j=1/2s^2$ and describes how well-ordered the layers are. The centre of the Gaussian distribution $a_j$ corresponds to the position of the $j$-th layer along the director. 
The distance between two consecutive layers is then calculated as $\lambda_j=a_{j+1}-a_j$. Beyond the anti-ferroelectric transition, the values of $\lambda_j$ with odd $j$ correspond to the distance between layers in which there are more attractive tips than tails facing each other. For this reason, the average value $\lambda_\textrm{odd}$ is expected to be lower than $\lambda_\textrm{even}$ beyond the anti-ferroelectric transition; $\lambda_\textrm{even}$ is then the average over the values $\lambda_j$ with even $j$ and over the equilibrium configurations for a given pressure. 

\emph{Smectic ordering potential - }The smectic ordering potential $\Delta{U(z)}$ describes the molecular field experienced by each individual particle in the smectic A phase. It is obtained from the distribution of the particles' centres of mass along the $z$-direction, which corresponds approximately to the director $\mathbf{n}$, $ \rho(z)$. 
More specifically, the relation $\rho(z)\propto{e}^{-\Delta{U(z)}/k_\mathrm{B}T}$ holds by virtue of assuming a Boltzmann distribution. We fit this relation to the simulation data to extract $\Delta{U(z)}$.
From the smectic ordering potential, we compare between various attraction strengths and volume fractions how well-defined the layers are and how difficult it is for a particle to hop from layer to layer. We quantify it by measuring the difference between the highest and lowest value of the smectic ordering potential $\Delta{U}$, which defines the height of the smectic ordering potential $h_{\Delta{U}}$, for several values of the attraction strength and volume fraction. We also estimate the width of the smectic ordering potential $w_{\Delta{U}}$ by calculating the full width at half maximum of $ \rho(z)$, using the standard deviation $s$ of the gaussian function described in the previous paragraph, giving $w_{\Delta{U}}=2\sqrt{2\ln{2}}s$.

\section{Phase behaviour}
\label{sec:phase}
\begin{figure*}[ht]
\includegraphics[scale=0.45]{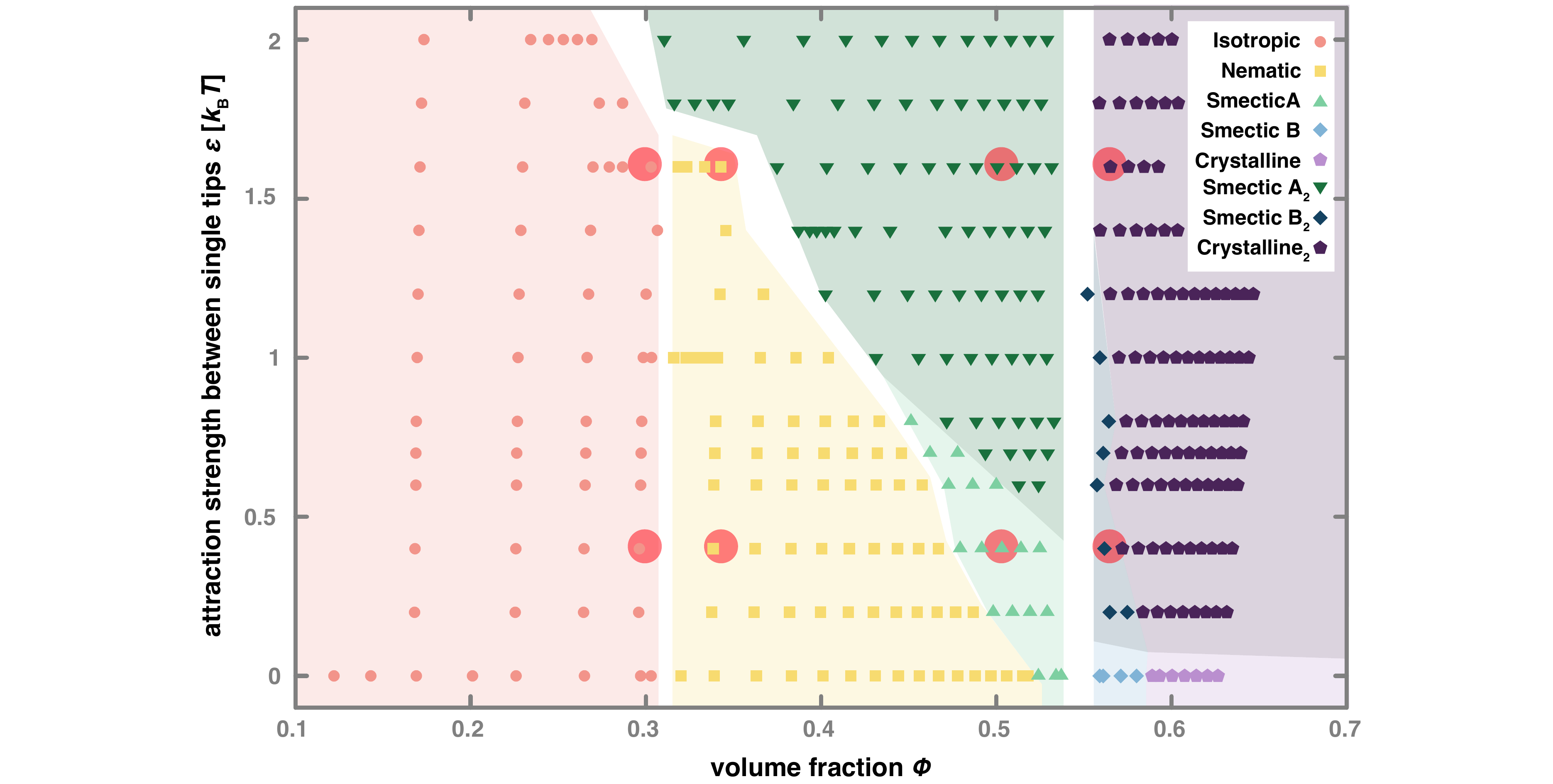}%
\caption{\label{imgPD}Calculated phase diagram of repulsive, rod-like particles that have a single attractive tip as a function of the attraction strength between the end tips (in units of thermal energy) and the volume fraction $\phi$. The particles have a base aspect ratio $L_0/D = 10.77$ and flexibility of $L/L_P = 0.3$. (See main text.) The following phases are identified: isotropic (orange circle), nematic (yellow square), smectic A (green triangle up), smectic A$_2$ (dark green triangle down), smectic B (blue diamond), smectic B$_2$ (dark blue diamond), crystalline (purple pentagon) and crystalline$_2$ (dark purple pentagon).  Snapshots of the data points highlighted by red circles in the phase diagram are presented in Figure \ref{imgSnap}.}%
 \end{figure*}
   
We focus attention on the phase behaviour of semi-flexible rod-like particles that have an aspect ratio of $L_0/D=10.77$ and a flexibility of $L_0/L_p=0.3$. 
The flexibility of our model particles matches that of the M13 virus investigated in Reference \cite{PatchyRef} albeit that the aspect ratio of our particles is considerably shorter by a factor of 10. 
Actually, the effective aspect ratio of the viruses, accounting for the electric double layer of the M13 particles, is about three times our aspect ratio. 
We choose to simulate shorter particles, because it enables us to investigate them for a wide range of concentrations, keeping the same number of particles in our simulation box and the total simulation time. 
In general, as the aspect ratio of the particles is reduced, the phase transitions of their suspensions are shifted to larger volume fractions because the excluded volume interaction is less anisotropic. Still, both experimental and simulation particle models support the same phases except for the columnar phase, which has not yet been observed in particle-based simulations involving monodisperse particles \cite{Wilson1994,Bolhuis1997,Cinacchi2008,Braaf2017}. Therefore, we are able to compare them in what is our main interest in this paper: the increase of the stability of the smectic A phase at the expense of the nematic phase. 

In the phase diagram presented in Figure \ref{imgPD}, the phase sequence at the zero attraction strength, $\epsilon=0\,k_\mathrm{B}T$, is taken from our previous work \cite{Braaf2017}. In this case, the particles are purely repulsive and have the same total number of beads as the particles with an attractive tip. 
Note that as they are not equipped with the attractive bead these rods are not polar. For these purely repulsive particles, the phase sequence consists of the following phases from lower to higher volume fractions: isotropic, nematic, smectic A, smectic B, and crystalline phases. 
From these simulations, we find that the isotropic-to-nematic and the smectic A-to-smectic B phase transitions seem to be of the first order.
The smectic B-to-crystalline phase transition appears to be continuous, while the nematic-to-smectic A phase transition is either continuous or weakly first order.
In the recent work of Milchev \emph{et al.}, in which very large-scale simulations of semi-flexible particles are performed, the latter transition is continuous \cite{Milchev2019}.
For the model particles described in the methods section, we vary the depth of the attraction well $\epsilon$ of the Lennard Jonnes potential between the beads representing the functionalised tips of the viruses, that is, the \emph{attraction strength}, from 0.2 to 2\,$k_\mathrm{B}T$. The calculated volume fraction is within the range of approximately 0.1 to 0.7. In order to obtain more resolution in volume fraction near the isotropic-nematic transition, we perform additional simulations for the attraction strengths 1 and $1.6\,k_\mathrm{B}T$ near the transition. We do likewise for the nematic-smectic A and isotropic-smectic A phase transitions for the attraction strengths $1.4$ and $1.8\,k_\mathrm{B}T$. We separate our discussion of the phase diagram in the next two paragraphs, related to the low- and a medium-energy regime of the single-end, attractive interaction on account of the qualitative difference in behaviour.

\emph{Lower attraction strengths - }For values of the attraction strength up to about $1\,k_\mathrm{B}T$, our simulations show that there is no significant change in the isotropic-to-nematic phase transition. 
Comparing the sequences at 0 and $1\,k_\mathrm{B}T$, which is the one with more resolution in volume fraction in this regime of attraction strengths, we find that the largest and lowest volume fractions in which these phases are stable in our simulations (corresponding to the coexistence concentrations) coincide. 
For this reason, the isotropic-to-nematic phase transition remains first order.
We find that the nematic phase is destabilised in favour of the smectic A phase and that this phase transition becomes more strongly first order as the strength of the attraction interaction is increased.
These findings are consistent with the recent experimental observations on aqueous suspensions of M13 virus \cite{PatchyRef}. 
Our most remarkable finding is an anti-ferroelectric phase transition within the smectic A phase.
We identify the new smectic A$_2$ phase, characterised by a bi-layer type structure, in our phase diagram at $0.4\,k_\mathrm{B}T$ and higher attraction strengths, depending on the density. Section \ref{sec:mic} provides more details about the anti-ferroelectric transition, which in our simulations is continuous. 
There are also anti-ferroelectric smectic B and crystalline phases. 
In fact, even at the lowest attraction strength investigated, we find only the anti-ferroelectric smectic B$_2$ and crystalline$_2$ phase, suggesting that only tiny interaction energies are required to stabilise these. 
We did not attempt to pinpoint at what low value of $\epsilon$ the transition happens for these phases.
As for the other phase transitions, we find that there is no significant change in the smectic A or smectic A$_2$ to smectic B$_2$ phase transition. 
Furthermore, the smectic B$_2$ phase destabilises with increasing value of $\epsilon$ and the crystalline$_2$ phase becomes more stable at lower volume fractions. We understand that the suppression of the smectic B$_2$ phase is due to the increased ordering of particles due to the sticky end groups, as we shall see below when discussing the changes in the microstructure of the phases with increasing attraction between the tips. 

\emph{Higher attraction strengths - }In order to investigate the effect of larger attraction energies between the end groups, we performed simulations for attraction strengths up to $2\,k_\mathrm{B}T$. The phase sequence at $1.2\,k_\mathrm{B}T$ follows the trends described in the previous paragraph. 
For this reason, the coexistence concentrations (volume fractions) of the nematic and smectic A$_2$ phases are even lower and the difference between them (the phase gap) becomes larger.
For stronger attraction, the smectic B$_2$ disappears, and as a result the smectic A$_2$-to-smectic B$_2$ phase transition is replaced by a smectic A$_2$-to-crystalline$_2$ transition. This transition seems to be independent of the attraction strength and the phases have similar coexistence volume fractions as the transition at lower values of the attraction strength. 
Between attraction strengths 1.6 and $1.8\,k_\mathrm{B}T$, we find that the nematic phase is completely suppressed in favour of the smectic A$_2$, thus also affecting the stability of the isotropic phase. Therefore, the phase sequence at the highest attraction strengths 1.8 and $2\,k_\mathrm{B}T$ consist of only three phases: isotropic, smectic A$_2$, and crystalline$_2$. 
From these results, we find that the isotropic phase may also be further destabilised with increasing attraction strength between the tips. From the simulations at increased resolution, we are able to conclude that the isotropic-nematic phase gap in volume fraction is about 0.02 for the attraction strength of $1\,k_\mathrm{B}T$ and only slightly larger for $1.6\,k_\mathrm{B}T$, corresponding to approximately 0.03. Therefore, the order of this phase transition does not seem to be strongly affected by the attractive end aggregation. On the other hand, the nematic-smectic A$_2$, as previously discussed, and isotropic-smectic A phase transitions become more strongly first order with an increase of the attraction strength, as we find from the more detailed sequences at the attraction strengths of $1.4$ and $1.8\,k_\mathrm{B}T$. 

Next, we discuss in more detail the microscopic structure of the various phases.

\section{Microstructure of the phases}
\label{sec:mic}
In this section, we present the most salient features of the microstructure of the various phases, focusing in particular on those of the smectic A and smectic A$_2$ phases in order to better understand what drives the anti-ferroelectric phase transition. 
First, we discuss the qualitative and quantitative changes in the aggregation of the particles with attractive tips. For this purpose, we investigate snapshots as well as the aggregation statistics of the particles in the various phases. 
Second, we discuss the behaviour of the interlayer distance for the smectic and crystalline phases, as well as their anti-ferroelectric version based on results from the two different analysis procedures described in Section \ref{sec:met}.
Next, we provide more details about the order of the anti-ferroelectric phase transition, presenting the order parameter used for the classification of the smectic A and smectic A$_2$ phases. 
Finally, we analyse the stability of the smectic A and smectic A$_2$ at a constant volume fraction, using the smectic ordering potential for various attraction strengths. For the different phases, we identify the following features.

\begin{figure}[ht]
\includegraphics[scale=.27]{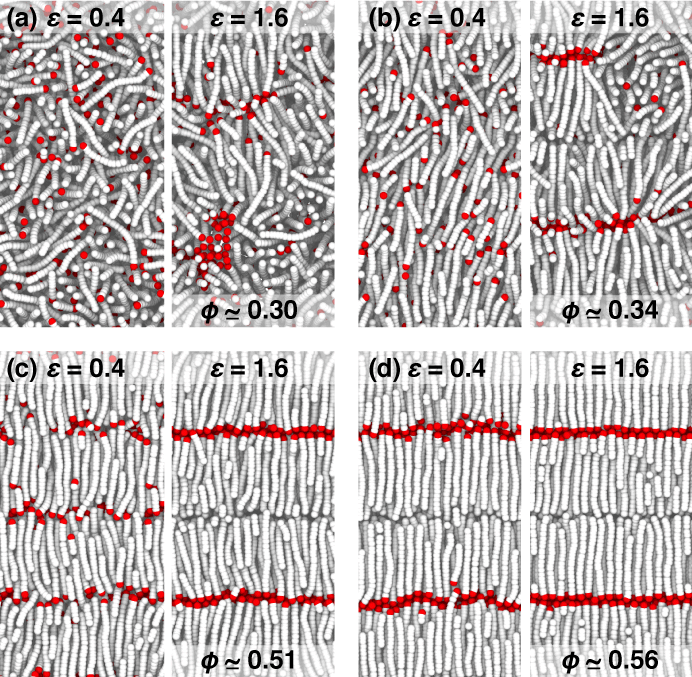}%
\caption{\label{imgSnap}Snapshots of the simulations at approximately constant volume fraction $\phi$ in the (a) isotropic phase at $\phi\simeq0.30$, (b) nematic phase at $\phi\simeq0.38$, (c) smectic A and smectic A$_2$ phases at $\phi\simeq0.51$, (d) smectic B (left) and crystal phases (right) at $\phi\simeq0.56$ obtained. From left to right, pair of snapshots for each value of the volume fraction are given for attractions strengths of 0.4 and $1.6\,k_\mathrm{B}T$. The corresponding data points are highlighted by red circles in the phase diagram in Figure \ref{imgPD}.}%
 \end{figure}

\emph{Clustering of particles - }
In the next paragraphs, we describe how the aggregation of elongated particles is influenced by the strength of the attractive tips in the various phases. 
The snapshots in Figure \ref{imgSnap} represent the aggregation patterns at approximately constant volume fraction $\phi$ in the isotropic phase at $\phi\sim0.30$ (a), nematic phase at $\phi\sim0.38$ (b), smectic A and smectic A$_2$ phases at $\phi\sim0.51$ (c), smectic B and crystal phases at $\phi\sim0.56$ (d). Left to right, each pair of snapshots shows examples of configurations of particles with attraction strengths of 0.4 and $1.6\,k_\mathrm{B}T$. 
The snapshots show that the aggregation pattern of the tips is clearly distinct, depending on the attraction strength for particles in the same phase and at the same volume fraction, as we describe in detail in the paragraph below.

From our phase diagram in Figure \ref{imgPD}, we find that the isotropic-to-nematic transition is only affected for $\epsilon\geqslant1.6\,k_\mathrm{B}T$. At this attraction strength, lamellar, disk- or inverted-micelle-like structures are formed. 
These two different aggregation patterns are represented in the snapshots on the right in Figure \ref{imgSnap} (a) and (b). In Figure \ref{imgSnap} (a), we find that, overall, the particles have random orientations, as expected for the isotropic phase. The particles aggregated in lamellar structures have similar orientation but the structures themselves have diverse orientations. In Figure \ref{imgSnap} (b), at attraction strenght $1.6\,k_\mathrm{B}T$ (right), we note that the alignment of the particles in the nematic phase results in a structure that resembles a highly disordered smectic phase. In view of that, we conclude that the formation of such lamellar structures must be the reason that there is a suppression of the nematic phase in favour of the smectic phase. 
In the snapshots of Figure \ref{imgSnap} (c), we compare how particles are organised along the director in the smectic A (left) and the smectic A$_2$ (right) phases at the same volume fraction. 
We find that the layers become more well-defined with increasing attraction strength. 
This is confirmed comparing the snapshots of Figure \ref{imgSnap} (d) for the smectic B$_2$ and crystalline$_2$ phases.
The attraction between the tips increases the degree of order of the particles, reducing the stability range of the smectic B$_2$ phase in favour of the crystalline phase.

\begin{figure}[ht]
\includegraphics[scale=.34]{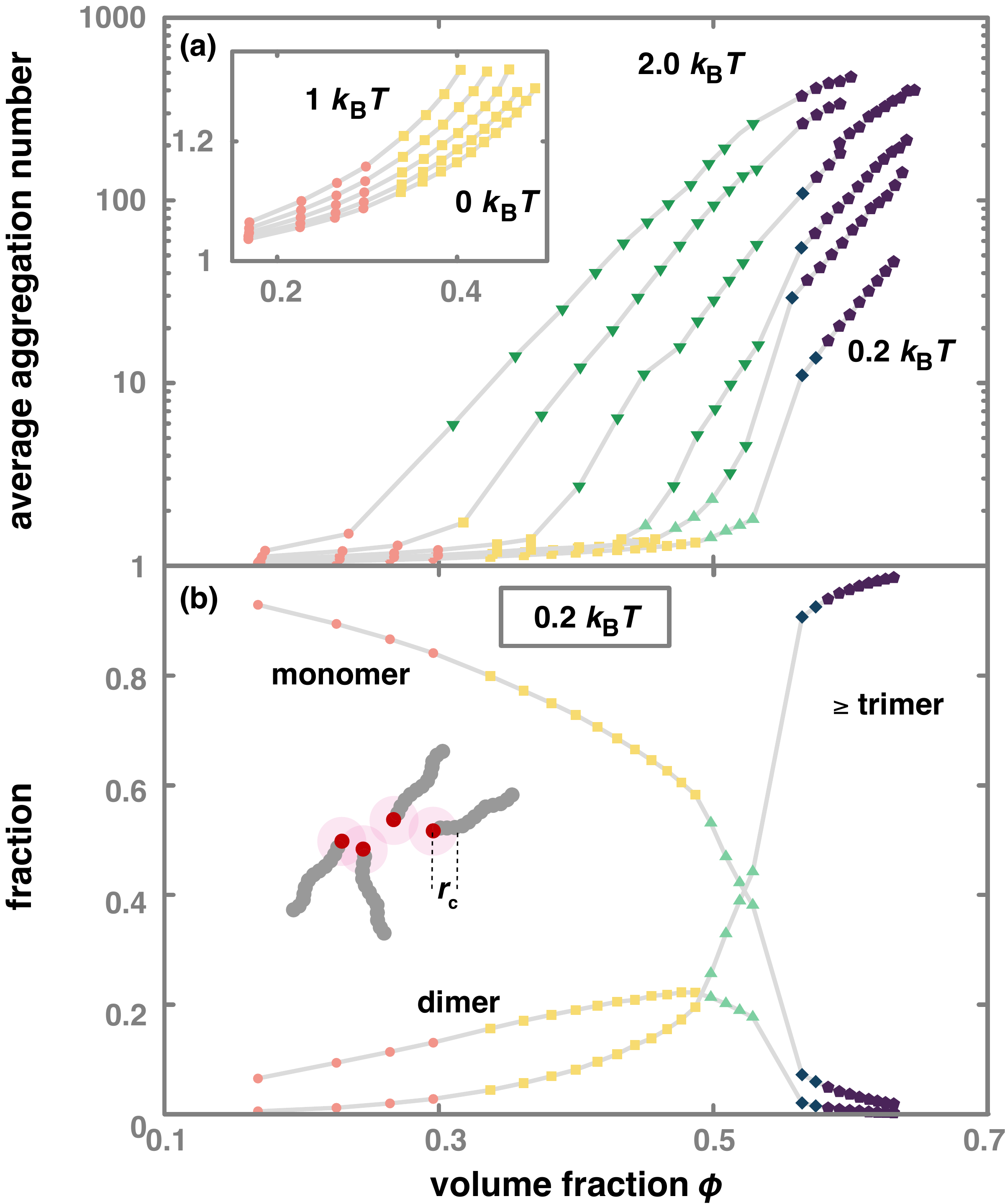}%
\caption{\label{imgCluster10}(a) Average aggregate size for various attraction strengths between the end groups as a function of the volume fraction of particles. Attraction strengths 0.2, 0.6, 0.8, 1.2, 1.6, and 2.0$\,k_\mathrm{B}T$ from right to left. The inset is an enlarged view of the graph for lower volume fractions for attractive strength equally spaced between 0 and 1$\,k_\mathrm{B}T$. (b) Fraction of monomers, dimers, trimers or larger aggregates, which are connected via the attractive ends for attraction strength $0.2\,k_\mathrm{B}T$. The particles have aspect ratio $L_0/D = 10.77$ and a flexibility of $L_0/L_P = 0.3$. The following phases are identified: isotropic (orange circle), nematic (yellow square), smectic A (green triangle up), smectic A$_2$ (dark green triangle down), smectic B$_2$ (dark blue diamond), and crystalline$_2$ (dark purple pentagon).}%
 \end{figure}
 
We compare these patterns with the corresponding aggregation statistics as a function of the volume fraction in Figure \ref{imgCluster10}, where we present (a) the average tip aggregation number for various attraction strengths and (b) the fraction of particles in monomers, dimers, and trimers or larger aggregates for an attraction strength of 0.2\,$k_\mathrm{B}T$. 
Figure \ref{imgCluster10} (a) shows, as expected, that the average aggregation number increases with increasing attraction strength and with increasing concentration due to mass action \cite{Pashley2004}. The microstructure therefore changes even in the isotropic phase, although, surprisingly, the isotropic-nematic phase transition is not affected at all. 
As we infer from the inset in Figure \ref{imgCluster10}, the tip clustering is weak at attraction strengths up to $1\,k_\mathrm{B}T$ in both the isotropic and the nematic phases. 
For these phases and attraction strengths, the aggregation numbers remain modest even though growth is stronger than a linear increase with the volume fraction. 
Actually, in these phases the average aggregation number is not larger than 1.3, which means that the tips are mostly monomers. 
Figure \ref{imgCluster10} (b) confirms this: in both the isotropic and the nematic phases, the fraction of monomers and dimers predominate. The fraction of trimers or larger aggregates surpasses the fraction of dimers only in the smectic A phase and then the fraction of monomers in the smectic B$_2$ phase. 
This indicates that larger aggregates are formed due to the inherent structure of the phase rather than due to the attraction strength alone. 
Indeed, part of the clustering is due to change in the contact value of the pair distribution function with the increase of pressure \cite{Henderson1992}.
For this reason, there is aggregation of tips even for $\epsilon=0\,k_\mathrm{B}T$ and the aggregation becomes much more prominent in the smectic and crystalline phases, and their anti-ferroelectric versions on account of the strongly increased pressures. 
Note the abrupt increase in size in going from the nematic to the smectic A or smectic A$_2$ phase in Figure \ref{imgCluster10} (a).  
 
\begin{figure}[ht]
 \includegraphics[scale=.34]{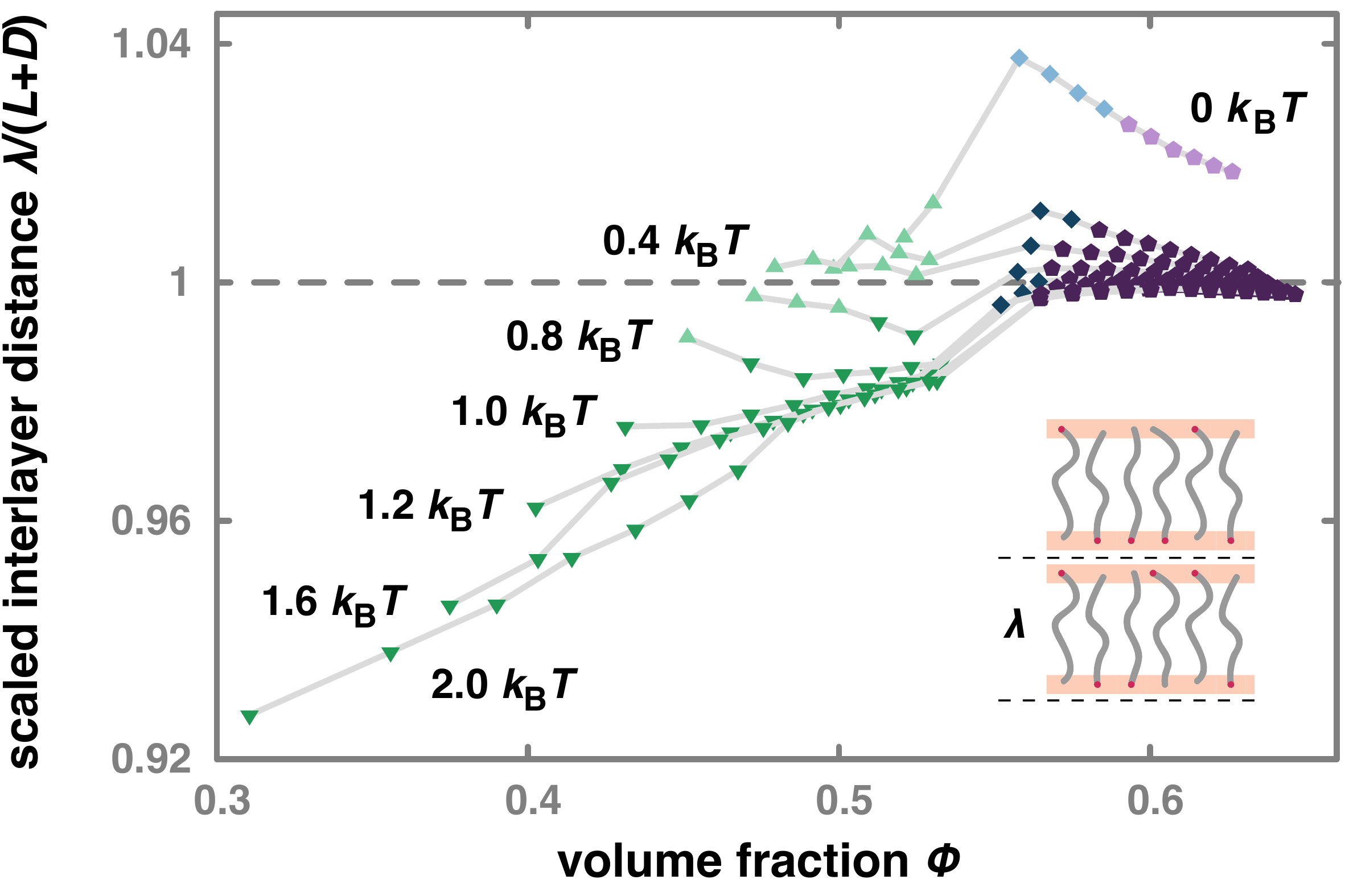}%
 \caption{\label{plotX} Scaled interlayer distance $\lambda/(L+D)$ of repulsive, rod-like particles with a single attractive tip as a function of the volume fraction $\phi$ for attraction strengths between the end groups ranging from 0 to $2\,k_\mathrm{B}T$ from top to bottom in the smectic A (green triangle up), smectic A$_2$ (dark green triangle down), smectic B (blue diamond), smectic B$_2$ (dark blue diamond), crystalline (purple pentagon) and crystalline$_2$ (dark purple pentagon) phases. Particles have aspect ratio $L_0/D = 10.77$ and a flexibility of $L_0/L_P = 0.3$. The inset is an illustration representing the interlayer distance $\lambda$.}%
 \end{figure}

\begin{figure}[ht]
 \includegraphics[scale=.34]{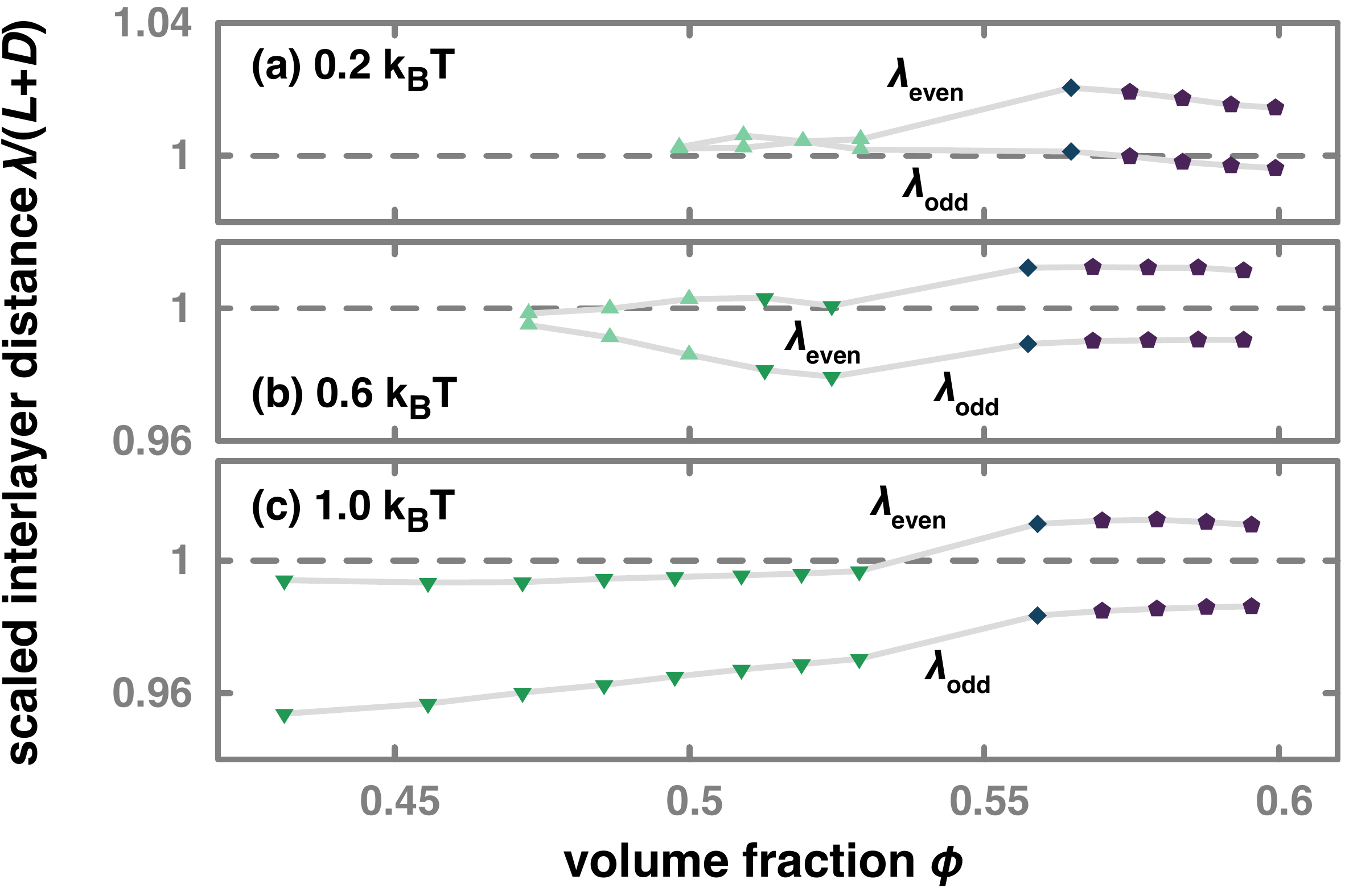}%
 \caption{\label{plotXb}Average interlayer distance relative to the average rod length $\lambda_\mathrm{odd}/(L+D)$ for odd and $\lambda_\mathrm{even}/(L+D)$ for even interlayer distances  of repulsive, rod-like particles that have an attractive tip in the smectic A (green triangle up), smectic A$_2$ (dark green triangle down), smectic B$_2$ (dark blue diamond), and crystalline$_2$ (dark purple pentagon) phases for attraction strengths of (a) $0.2\,k_\mathrm{B}T$, (b) $0.6\,k_\mathrm{B}T$, and (c) $1.0\,k_\mathrm{B}T$. Particles have aspect ratio $L_0/D = 10.77$ and flexibility of $L/L_P = 0.3$. }%
 \end{figure}

\emph{Interlayer distance -} 
The interlayer distance $\lambda$ corresponds to the average distance between the centre of masses of consecutive layers. 
This quantity comprises the average layer size added to the average interlayer gap, as represented in the inset of Figure \ref{plotX}, and depends on the characteristics and interactions of the particles.
As described in the methods section, we apply two different analysis procedures to our data in order to investigate the interlayer distance. We refer to the results relative to the standard procedure as the \emph{averaged} interlayer distance and to the results relative to the second procedure as the \emph{differentiated} interlayer distance. 
The main difference is that in the second procedure we differentiate between odd and even layer numbers. 
In Figure \ref{plotX}, we present the averaged interlayer distance $\lambda$ relative to the average particle length $L+D$, as a function of the volume fraction $\phi$ for attraction strengths between the end groups ranging in strength from 0 to $2\,k_\mathrm{B}T$. In Figure \ref{plotXb}, we present the values of the differentiated interlayer distance for even $\lambda_\mathrm{even}$ and odd $\lambda_\mathrm{odd}$ over the particle length $L+D$ as a function of the volume fraction $\phi$ for attraction strengths $0.2$, $0.6$, and $1.0\,k_\mathrm{B}T$. 
Note that the length $L$ is not the bare length but the actual, measured value, somewhat compressed by the ambient pressure.

In Figure \ref{plotX}, we find that the scaled interlayer distance $\lambda/(L+D)$ exhibits a rich behaviour depending on the attraction strength between the tips and on the state of aggregation. In the smectic A phase and at attraction strengths larger than or equal to $1\,k_\mathrm{B}T$, the scaled interlayer distance increases with the volume fraction. 
Note that these values are smaller than unity, which means that the layers slightly interdigitate. 
This interpenetration lowers the interaction energy because it allows for a larger surface contact between the tips, which consist of an exposed attractive hemispherical cap.
Between 0.8 and  $0.6\,k_\mathrm{B}T$, the interlayer distance seems to decrease with increasing volume fraction. Interestingly, it is for these strengths that we identify both the smectic A and the smectic A$_2$ phase. 
For lower attraction strengths, the scaled interlayer distance is slightly larger than unity, indicating that, in this case, layers are nearly touching each other. Nevertheless, the dependence on the volume fraction is not obvious. 
Overall, the scaled interlayer distance seems to decrease with increasing attraction strength between tips. This effect is unambiguous if we consider volume fractions lower than 0.45. 
In other words, at a fixed volume fraction, layers interpenetrate more as the attraction strength between tips increases.  
As a result, the inlayer density is expected to be smaller. 
In the smectic B and the crystalline phases or the smectic B$_2$ and the crystalline$_2$ phases, we note that the scaled interlayer distance decreases with volume fraction and that the value converges to unity with increasing volume fraction for all attraction strengths.

Figure \ref{plotX} clarifies several features of our system but hides the distinction between layers present in the anti-ferroelectric phases. In Figure \ref{plotXb}, which presents results from the differentiated analysis, there is the distinction between odd and even interlayers at three values of the attraction strength.  
At the first value of $0.2\,k_\mathrm{B}T$, represented in (a), we find that the interlayer distance is approximately the same for even and odd interlayers in the smectic A phase.
As expected, their values are also approximately equivalent to the averaged layer thickness, once the layers are nearly indistinct in this phase. 
Nevertheless, in the smectic B$_2$ and crystalline$_2$ phases, the values for the even and odd interlayer distance, which respectively contain the smaller and larger fraction of attractive tips, are distinct. 
The former is larger than unity and the latter is approximately unity. 
These values slightly decrease with the volume fraction. 
In Figure \ref{plotXb} (b), the results are for the attraction strength of $0.6\,k_\mathrm{B}T$. 
In the smectic A phase, we find that for that case the values of the odd and even interlayer distances are approximately equal only at the lowest volume fraction, and that they become distinct as the volume fraction increases. 
Notice that the even interlayer distances are larger than the odd ones, because even though the anti-ferroelectric order parameter $\delta$ is different from zero, it is smaller than 1/2 and hence the phase is not classified as smectic A$_2$.
The even interlayer distance is approximately constant and equal to unity, while the odd interlayer distance becomes smaller, meaning that the layers in these interfaces are also interpenetrating. 
In the smectic A$_2$ phase, the even interlayer distance follows the same trend as in the smectic A phase, while the odd interlayer is slight smaller but the dependence on volume fraction is not clear due to the lack of data points available in this phase. In the smectic B$_2$ and crystalline$_2$ phases, it seems that even and odd interlayer distances do not depend on the concentration. 
Their values are slightly above and below unity, respectively. We find the same for an attraction strength of $1.6\,k_\mathrm{B}T$, as we see in Figure \ref{plotXb} (c). 
For the smectic A$_2$ phase, we find that the values are rather different and that, while the trend for even interlayer distance remains as described before, the odd interlayer distance, which is smaller than unity, tends to slightly increase with increasing volume fraction. This effect is probably due to the lower in-layer density of particles at lower volume fractions, which offers more space for the particles to interpenetrate. 

\begin{figure}[htp]
 \includegraphics[scale=.34]{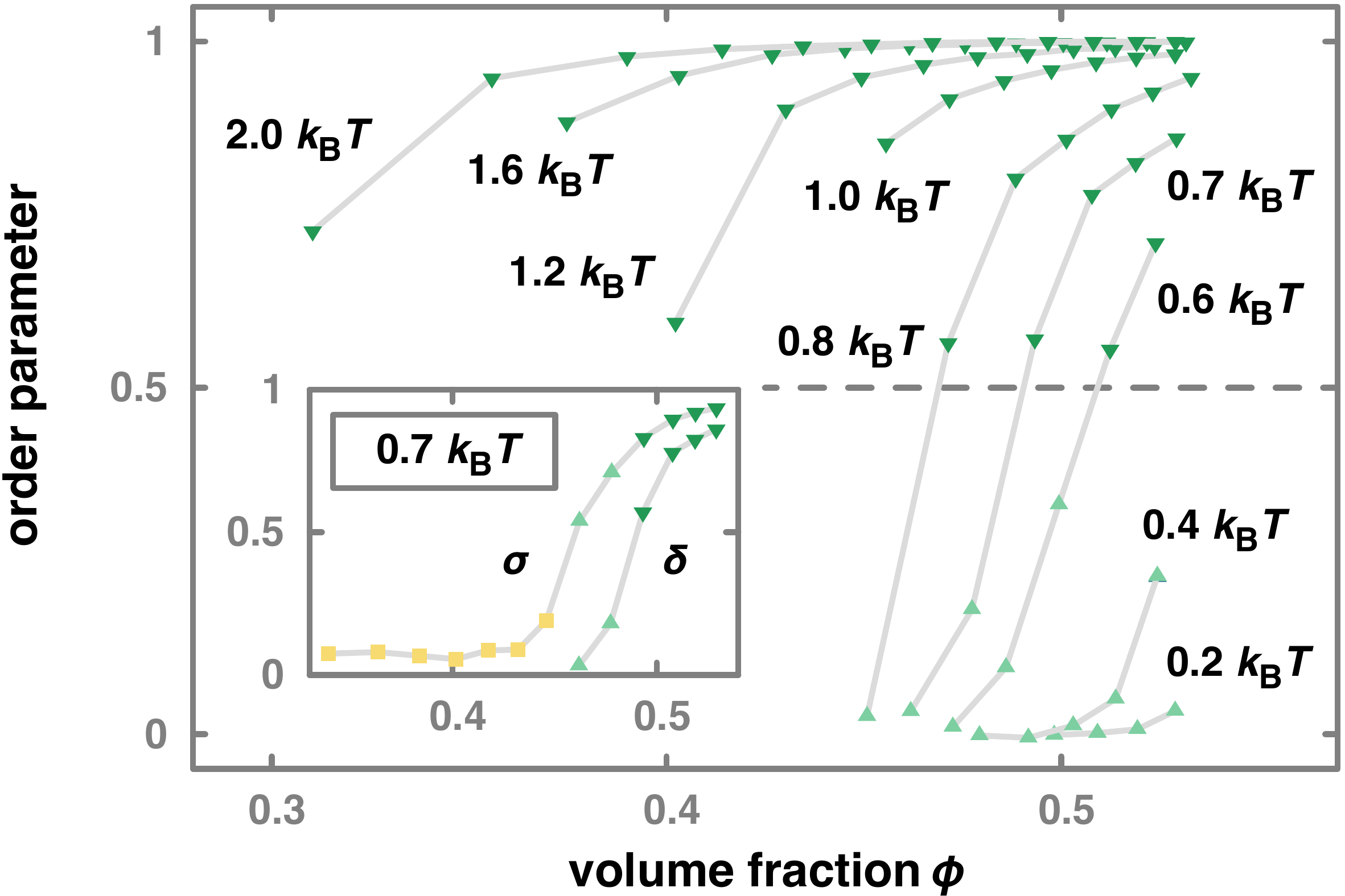}%
 \caption{\label{imgAnti}Anti-ferroelectric order parameter $\delta$ of repulsive, rod-like particles that have a single attractive tip as a function of volume fraction $\phi$ for various values of the attraction strength between the attractive tips. In the inset we compare the anti-ferroelectric $\delta$ and smectic $\sigma$ order parameters for the attraction strength of $0.7\,k_\mathrm{B}T$. Particles have aspect ratio $L_0/D = 10.77$ and flexibility of $L_0/L_P = 0.3$. The following phases are identified: nematic (yellow square), smectic A (green triangle up), and smectic A$_2$ (dark green triangle down) phases.}
\end{figure}
 
\emph{Anti-ferroelectric phase transition - }The anti-ferroelectric phase transition in the phase diagram is linked with the polarity of our single-ended attractive elongated particles. We identify this phase transition using the anti-ferroelectric order parameter $\delta$, defined in Section \ref{sec:met}, and presented in Figure \ref{imgAnti} as a function of the volume fraction $\phi$ for various values of the attraction strength. Its value continuously increases with increasing volume fraction, which indicates a second order phase transition. 
As previously mentioned, we are not able to pinpoint the anti-ferroelectric phase transition in the smectic B and crystalline phases: we find the smectic B$_2$ and crystalline$_2$ even at the lowest attraction strength investigated ($0.2\,k_\mathrm{B}T$).
In the inset, the anti-ferroelectric order parameter $\delta$ is represented together with the smectic $\sigma$ order parameter as a function of concentration for attraction strength $0.7\,k_\mathrm{B}T$. 
The smectic order parameter itself does not provide a clear indication of the presence of a anti-ferroelectric transition although that it could be masked by our limited resolution in the volume fraction $\phi$. For a comparison with our simple Maier-Saupe-McMillan model, where we do see much more enhanced smectic ordering beyond the transition, we refer to the Appendix. 

\begin{figure}[ht]
 \includegraphics[scale=.34]{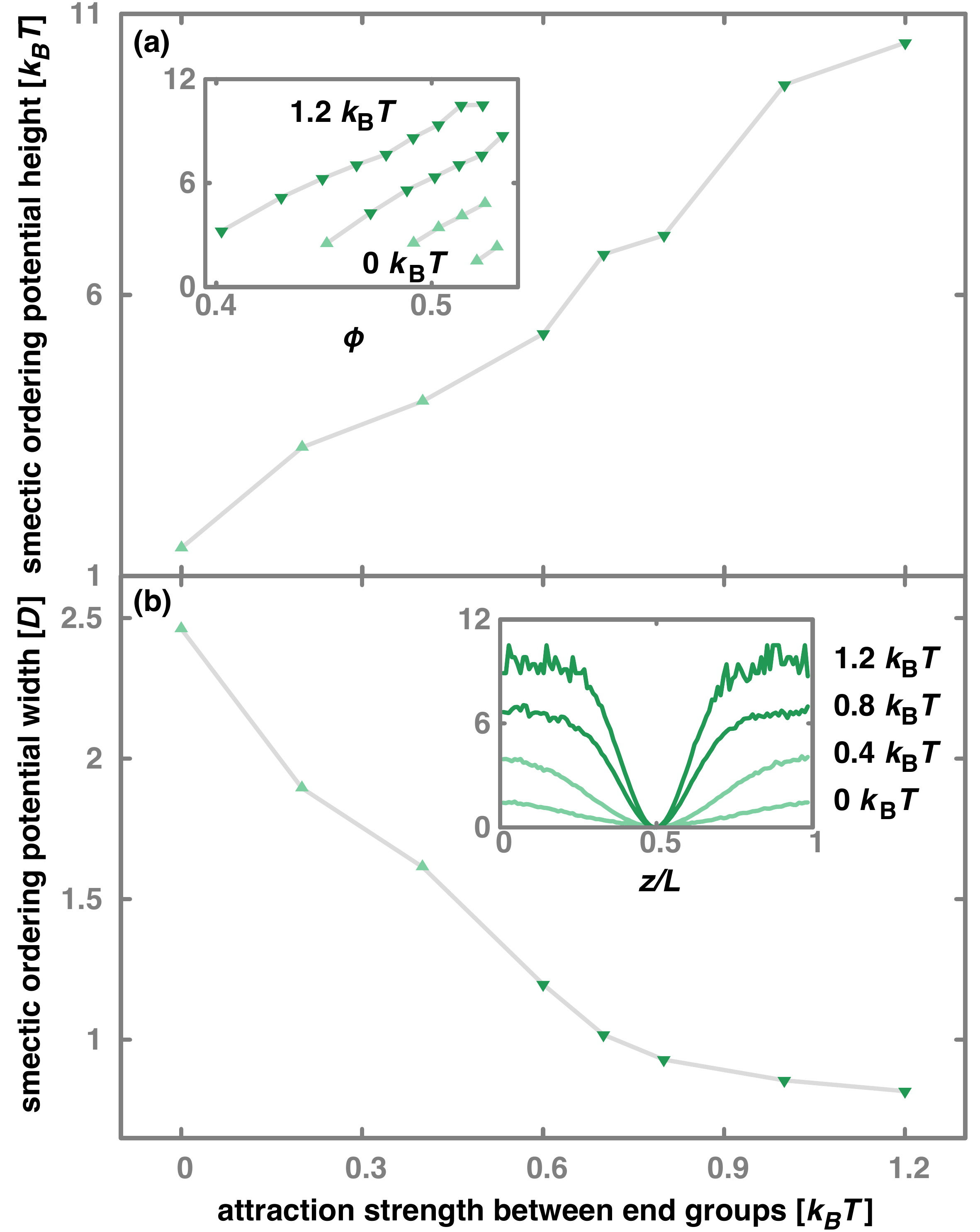}%
 \caption{\label{imgSmU}Smectic ordering potential of repulsive, rod-like particles with a single attractive tip. Particles have an aspect ratio of $L_0/D = 10.77$ and a flexibility of $L_0/L_P = 0.3$. (a) Height of the smectic ordering potential of repulsive as a function of the attraction strength between tips in the smectic A (green triangle up) or smectic A$_2$ (dark green triangle down) phases at a volume fraction of $\phi\cong0.52$. In the inset, there is the same height of the smectic ordering potential of repulsive as a function of the volume fraction for attraction strengths 0, 0.4, 0.8 and $1.2\,k_\mathrm{B}T$. (b) Width of the smectic ordering potential of repulsive as a function of the attraction strength between tips in units of $k_\mathrm{B}T$. In the inset, the smectic ordering potential $\Delta{U}(z)$ is presented as a function of the position along the director, normalised by the particle length $z/L$. The increase of the attraction strength drives the stabilisation of the smectic A and smectic A$_2$ phases.}%
 \end{figure}

\emph{Smectic ordering potential - } 
The smectic ordering potential $\Delta{U}(z)$ represents the molecular field that a particle experiences from the other particles within a smectic layer.
We calculate it for our simulations in the smectic A phase around the anti-ferroelectric phase transition, as represented in the top inset of Figure \ref{imgSmU}. 
From the inset, we find that both the height and the width changes as we increase the attraction strength at a constant volume fraction of $\phi=0.52$: the potential barrier increases and becomes narrower around the centre of the layer $z/L=0.5$. 
In other words, the smectic layers become increasingly ordered, as we also see from the snapshots in Figure \ref{imgSnap} (c). 
The smectic ordering potential seems to show more noise at higher attraction strengths, in particular in between the layers. This, in all likelihood, is due to the poor statistics: once the particle centres of masses are concentrated around the centre of the layer and few particles venture out in between the layers. 
From Figure \ref{imgSmU}, we find that the height of the smectic ordering potential linearly increases with the attraction strength at the volume fraction of $\phi=0.52$, and the trend does not seem to be changed crossing the anti-ferroelectric phase transition. 
The same is true if the volume fraction is increased at constant attraction strength, as it is shown in the top inset of Figure \ref{imgSmU}. Therefore, our particles are more strongly attached to a layered structure and, on top of that, the layer itself is more strongly confined due to the small amplitude in the fluctuations of the particle positions around the average. 
This molecular field of particles in the smectic phase as well as the smectic and anti-ferroelectric order parameters, presented in the previous paragraph, are also reasonably well-described by our simple Maier-Saupe-McMillan model (see Appendix). 

\section{Discussion and conclusion}
\label{sec:con}

The aim of the present work is to study by means of molecular dynamics simulations the influence that attractive interaction between one of the two ends of a collection of otherwise mutually repulsive, elongated particles have on their phase behaviour, using computer simulations. 
Our work shows that their phase behaviour, and the structure of the various liquid-crystalline phases, are strongly affected even by relatively weak interaction strengths on the order of the thermal energy. 
The phase behaviour has two striking features: (1) the formation of bi-layered anti-ferroelectric phases and (2) the large increase of stability of smectic A at the expense of the nematic phase, and even at the expense of the isotropic phase as we increase the attraction strength between the tips beyond about $1.6\,k_\mathrm{B}T$. The key factor is the interplay between the interaction energy, polarity of the particles, and the phase microstructure. The analysis of the microstructure reveals that the stability of the isotropic and nematic phases is affected by the aggregation of the tips, only if the particles align in response to the local increase in concentration. 
In the layered phases, the organisation of the particles with the attractive tips results in more strongly ordered microstructures even at very weak attraction energies. 

Our results concerning the shift of the nematic-to-smectic A phase transition to lower volume fractions, which also seems to become more strongly first order with an increase of the attraction strength, are supported by experimental evidence.
Despite the fact that the difference in the aspect ratio between simulated and experimental particles prevents us to make a quantitative comparison, we are able to qualitatively compare the smectic ordering potential of the experimental and computational model particles. We find for both of them larger values of the height and smaller values of the width as the attraction strength between tips increases. As expected, because the experimental model particles are larger in aspect ratio, their smectic potential height is also larger, due to the increase of stability of the smectic A phase with increasing aspect ratio \cite{Repula2018}. 
In our theory (Appendix \ref{apndx}), the shift of the nematic-to-smectic A phase transition is independent of the volume fraction. Nevertheless, the shift to lower volume fractions is captured if the nematic-to-smectic A$_2$ phase transition is considered. The height and width of the effective smectic ordering potential from our Maier-Saupe-McMillan theory does not emulate the smectic ordering potential from our simulations. This might be due to our choice of representing it as a cosine function instead of a Gaussian function.

In spite of the obvious limitations, our study does contribute to the understanding of how selective surface functionalisation of colloidal liquid crystals affects their self-organisation, by providing a systematic study of the stability and structure of these phases for a wide range of both volume fractions and attraction strengths. 
We show that incorporating a single (enthalpic) functionalised end in elongated colloidal particles gives rise to an even more complex and rich phase behaviour than for the purely repulsive ones. In the light of this, we suggest that agreement between purely repulsive models and experiment cannot be expected because residual attractive interactions, local or global, strongly influence the phase behaviour.  

\begin{acknowledgments}

This project has received funding from the European Union's Horizon 2020 research and innovation programme under the Marie Sklodowska-Curie Grant Agreement No 641839. We thank Thijs W. G. van der Heijden for the assistance in implementing the numerical integration of the equations presented in the Appendix. 
 
\end{acknowledgments}

\begin{appendix}

\section{The anti-ferroelectric phase transition within a Maier-Saupe-McMillan type theory}
\label{apndx}

Here, we describe a simplified model for the anti-ferroelectric phase transition for $N$ perfectly parallel, rod-like particles, equipped with a single tip attractive interaction. These particles are aligned along the $z$-axis, forming the nematic phase if they are uniformly distributed, or smectic phases if their positions are periodically distributed along the $z$-axis. The periodicity in the smectic phases is $d$, corresponding to the spacing between the layers. The single tip attraction is the key element for the existence of the smectic A$_2$ phase. In this bilayered phase, the distribution of the attractive tip has the periodicity $2d$.  

Our description is inspired by Maier-Saupe-McMillan theory\cite{Maier1959,McMillan1971}. We write the Helmholtz free energy $F$ in one layer as the sum of the Gibbs entropy, proportional to the position density distribution along the director of both the elongated particles $p(z)$ and the attractive tips $f(z)$, and energy terms proportional to the molecular field or smectic ordering potential, $\Delta{U_\mathrm{SmA}(z)}$, and the interaction energy between attractive tips, ${\Delta{U_\mathrm{SmA_2}(z)}}$,
\begin{equation*}
\begin{split}
\frac{\beta{F}}{N}= & \int_{-\frac{d}{2}}^{+\frac{d}{2}}dz\left[p(z)\ln{p(z)}+\frac{1}{2}p(z)\beta{\Delta{U_\mathrm{SmA}(z)}}\right] \\
& + \int_{-\frac{d}{2}}^{+\frac{d}{2}}dz\left[f(z)\ln{f(z)}+\frac{1}{2}f(z)\beta{\Delta{U_\mathrm{SmA_2}(z)}}\right].
\end{split}
\end{equation*}
The factor $1/2$ in the energy terms corrects for double counting: the distribution functions $f(z)$ and $p(z)$ are properly normalised. 

The energy term $\beta{\Delta{U_\mathrm{SmA}(z)}}$ drives the nematic-smectic A phase transition and it is expected to be proportional to the volume fraction $\phi\in[0,1]$ and the smectic ordering parameter $\sigma$, so
\begin{equation}
\beta{\Delta{U_\mathrm{SmA}(z)}}= -\phi\gamma\sigma\cos\left(\frac{2\pi{z}}{d}\right),
\label{eqEn}
\end{equation}
where $\gamma$ is an adjustable parameter and the cosine describes the periodic molecular field (of periodicity $d$). The smectic ordering parameter mentioned previously in this work is given by
\begin{equation}
\sigma=\int_{-\frac{d}{2}}^{+\frac{d}{2}}dz\,{p(z)}\cos\left(\frac{2\pi{z}}{d}\right).
\label{eqSigma}
\end{equation}

\begin{figure}[ht]
 \includegraphics[scale=.3]{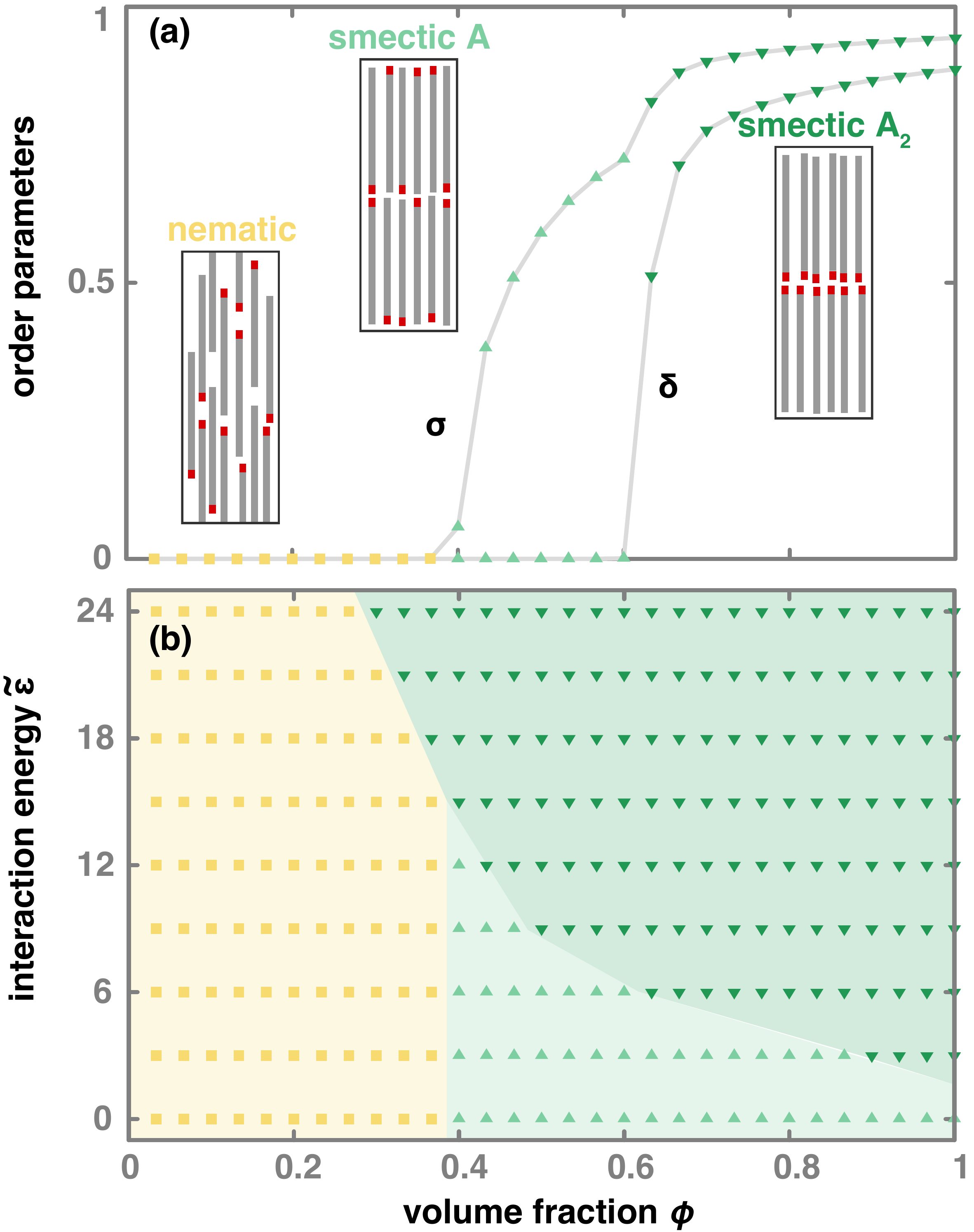}%
 \caption{\label{imgA}Theoretical approach to describe the anti-ferroelectric phase transition for perfectly parallel, hard rods that have a single attractive end. Phases identified are: nematic (yellow square), smectic A (green triangle up), and smectic A$_2$ (dark green triangle down). (a) Smectic and anti-ferroelectric order parameters numerically calculated for the attraction strength $\epsilon=6$ and the adjustable parameter $\gamma=5$, chosen to adjust the volume fraction interval for convenient viewing. (b) Phase diagram obtained from classification based on the order parameters for the same value of the adjustable parameter $\gamma=5$. We find that the nematic phase is destabilised in favour of the smectic A$_2$ phase and that the smectic A phase is suppressed at high enough attraction strength $\epsilon$.}%
 \end{figure}

Note that \ref{eqEn} is temperature invariant: it represents the hard-core nature of the interactions. The energy term $\beta{\Delta{U_\mathrm{SmA_2}(z)}}$ drives the anti-ferroelectric transition and it should arguably be also proportional to the volume fraction $\phi$ and some power of the smectic ordering parameter $\sigma$, as well as to the anti-ferroelectric order parameter $\delta$ itself. We put forward
\begin{equation*}
\beta{\Delta{U_\mathrm{SmA_2}(z)}}= -\phi\,{\sigma}^2\tilde{\epsilon}\,\delta\sin\left(\frac{2\pi{z}}{2d}\right),
\end{equation*}
where $\tilde{\epsilon}$ is a (dimensionless) measure for  the strength of the attraction between the tips and the sine function represents the periodicity $2d$ of the anti-ferroelectric state.
There is no obvious mapping of the sticking energy of our simulations and that of our model, although we would guess that $\tilde{\epsilon}\propto\epsilon$.
The anti-ferroelectric ordering parameter is given by  
\begin{equation}
\delta=\int_{-\frac{d}{2}}^{+\frac{d}{2}}dz\,{f(z)}\sin\left(\frac{2\pi{z}}{2d}\right).
\label{eqDelta}
\end{equation}

In equilibrium, the free energy must be functionally minimised, ${\delta\left(\beta{F}/N\right)}/{\delta{p}}=\lambda$ and ${\delta\left(\beta{F}/N\right)}/{\delta{f}}=\mu$, accounting for the normalisation conditions
\begin{equation*}
\int_{-\frac{d}{2}}^{+\frac{d}{2}}dz\,p(z)=\int_{-\frac{d}{2}}^{+\frac{d}{2}}dz\,f(z)=1,
\end{equation*}
which require us to introduce the Lagrange multipliers $\lambda$ and $\mu$. 

Making use of the normalisation of $p(z)$, we find that 
\begin{equation}
p(z)=\frac{\exp{\left[\sigma\phi(\gamma+\tilde{\epsilon}\delta^2)\cos\left(\frac{2\pi{z}}{d}\right)\right]}}{\int_{-\frac{d}{2}}^{+\frac{d}{2}}dz\exp{\left[\sigma\phi(\gamma+\tilde{\epsilon}\delta^2)\cos\left(\frac{2\pi{z}}{d}\right)\right]}}.
\label{eqP}
\end{equation}
Note that $\tilde{\epsilon}$ is an energy scaled to the thermal energy and hence temperature dependent. 
From thermodynamics, we have $\tilde{\epsilon}(T)=\tilde{\epsilon}(T_0)-\tilde{h}(T_0)(T-T_0)/T_0$, with $\tilde{h}$ a dimensionless enthalpy and $T_0$ a reference temperature. 
For hydrophobic interactions $\tilde{h}<0$ implying that the molecular field increases in strength with increasing temperature. 

\begin{figure}[ht]
 \includegraphics[scale=.3]{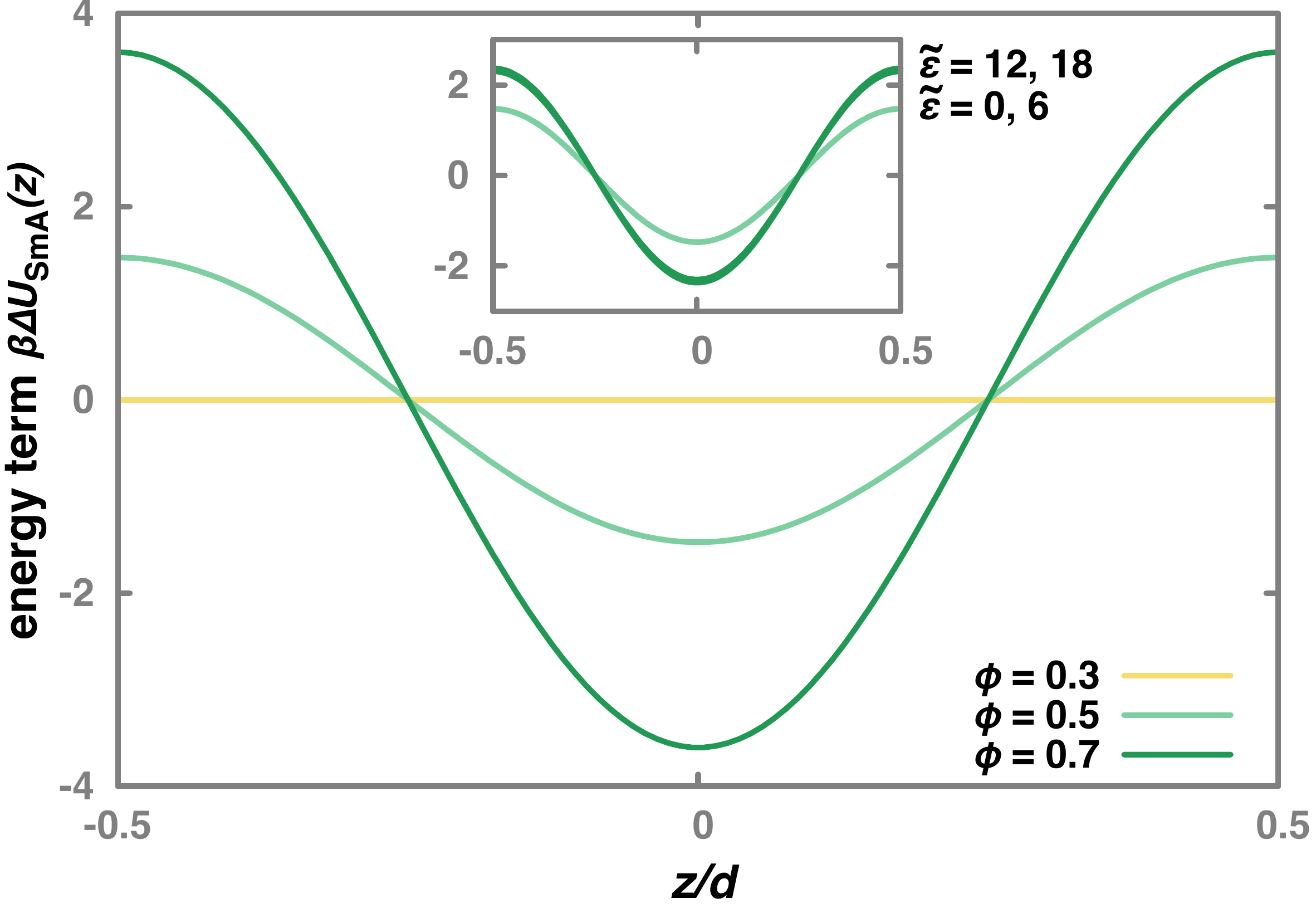}%
 \caption{\label{imgB}Energy term $\beta{\Delta{U_\mathrm{SmA_2}(z)}}$ or smectic ordering potential as a function of the position normalised by the layer thickness numerically calculated for the attraction strength $\epsilon=6$ and the adjustable parameter $\gamma=5$ in the nematic (at volume fraction $\phi=0.3$), smectic A (at $\phi=0.5$), and smectic A$_2$ (at $\phi=0.7$) phases. In the inset, the same energy term $\beta{\Delta{U_\mathrm{SmA_2}(z)}}$ calculated for the volume fraction $\phi=0.5$ and the adjustable parameter $\gamma=5$ in the smectic A (at attraction strengths $\tilde{\epsilon}=0$ and 6), and smectic A$_2$ (at $\tilde{\epsilon}=12$ and 18) phases. }%
 \end{figure}

We employ the same procedure for $f(z)$, and find 
\begin{equation}
f(z)=\frac{\exp{\left[\delta\phi\sigma^2\tilde{\epsilon}\sin\left(\frac{2\pi{z}}{2d}\right)\right]}}{\int_{-\frac{d}{2}}^{+\frac{d}{2}}dz\exp{\left[\delta\phi\sigma^2\tilde{\epsilon}\sin\left(\frac{2\pi{z}}{2d}\right)\right]}}.
\label{eqF}
\end{equation}
We recursively solve the coupled integral equations numerically for the smectic and the anti-ferroelectric order parameters, given by Equations \ref{eqSigma} and \ref{eqDelta}, in which we substitute the explicit expression for the density distribution of both the elongated particles $p(z)$ and the attractive tips $f(z)$, from Equations \ref{eqP} and \ref{eqF}, using the Mathematica software. 
We fix the value for adjustable parameter $\gamma=5$ in order to shift the transitions volume fractions to values close to where we find the transitions in our simulations. 
The initial values for the the smectic and the anti-ferroelectric order parameters is chosen to be 1. From each iteration, we obtain new estimates for $\sigma$ and $\delta$ that will be the input values to the next one. The volume fraction is fixed and increased from 0 to 1 in 30 steps. The sticking energy is also fixed at values between $\tilde{\epsilon}=0$ and 24, which we increased at steps of 3 in each run of our numerical calculations. The convergence criterion is that consecutive integration results differ less than 0.001 for at least 10 iterations. 

From the results obtained using the procedure described in the paragraph above, we calculate the values for the order parameters $\sigma$ and $\delta$, with which we classify the corresponding phase at the fixed set of sticking energy and volume fraction parameters. From this procedure, we obtain the phase diagram of perfectly parallel, hard rods that have a single attractive end. In Figure \ref{imgA}, we present in (a) the smectic $\sigma$ and anti-ferroelectric $\delta$ order parameters as a function of the volume fraction $\phi$ at sticking energy $\tilde{\epsilon}=6$ (and $\gamma=5$) and in (b) the phase diagram obtained from classification based on these order parameters.

 \begin{figure}[ht]
 \includegraphics[scale=.3]{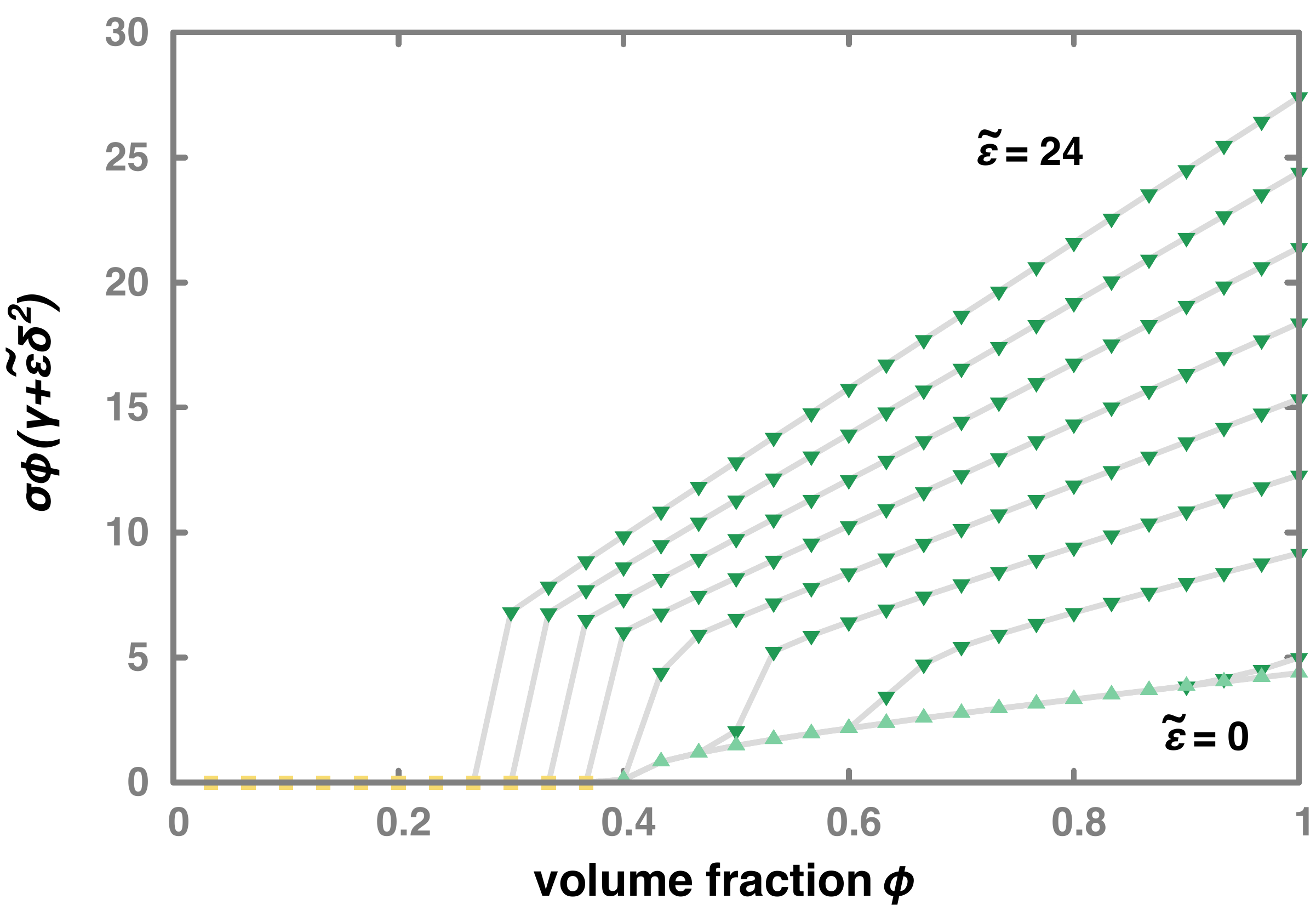}%
 \caption{\label{imgAextra1}Height of the effective smectic ordering potential is given by $\sigma \phi (\gamma + \tilde{\epsilon} \delta ^2)$ as a function of the volume fraction for the attraction strengths $\tilde{\epsilon}$ from 0 to 24 at the interval of 3. The following phases are identified: nematic (yellow square), smectic A (green triangle up), smectic A$_2$ (dark green triangle down) phases. The linear dependence of the amplitude of the smectic ordering potential with the volume fraction is also what we find in our simulations. See inset in Figure \ref{imgSmU}(a).}%
 \end{figure}
 
From Figure \ref{imgA} (a), we find that, unlike the corresponding order parameters obtained from our simulations, there is a shoulder in the smectic ordering parameter at the volume fraction where the anti-ferroelectric phase transition occurs. It indicates that the particles become more strongly ordered along the director as the rods transition to the smectic A$_2$ phase. 
The effect is not captured by the smectic order parameter in our simulations but this could be due to the limited resolution in our simulations. For a comparison, see the inset in Figure \ref{imgAnti} (a). From Figure \ref{imgA} (b), we find that our model captures the destabilisation of the nematic phase in favour of the smectic A$_2$ but the same is not true for the smectic A. In this case, the phase transition is independent of the sticking energy. Nevertheless, the destabilisation of the smectic A in favour of the smectic A$_2$ phase seems to represent what is seen in the simulations. From Figure \ref{imgB}, we find that the height is approximately the same in each phase, independently of the attraction strengths between tips, unlike what we find in the simulations. The same is true for the width of the smectic ordering parameter. It is independent of both the attraction strength and the volume fraction.
 
\begin{figure}[ht]
 \includegraphics[scale=.3]{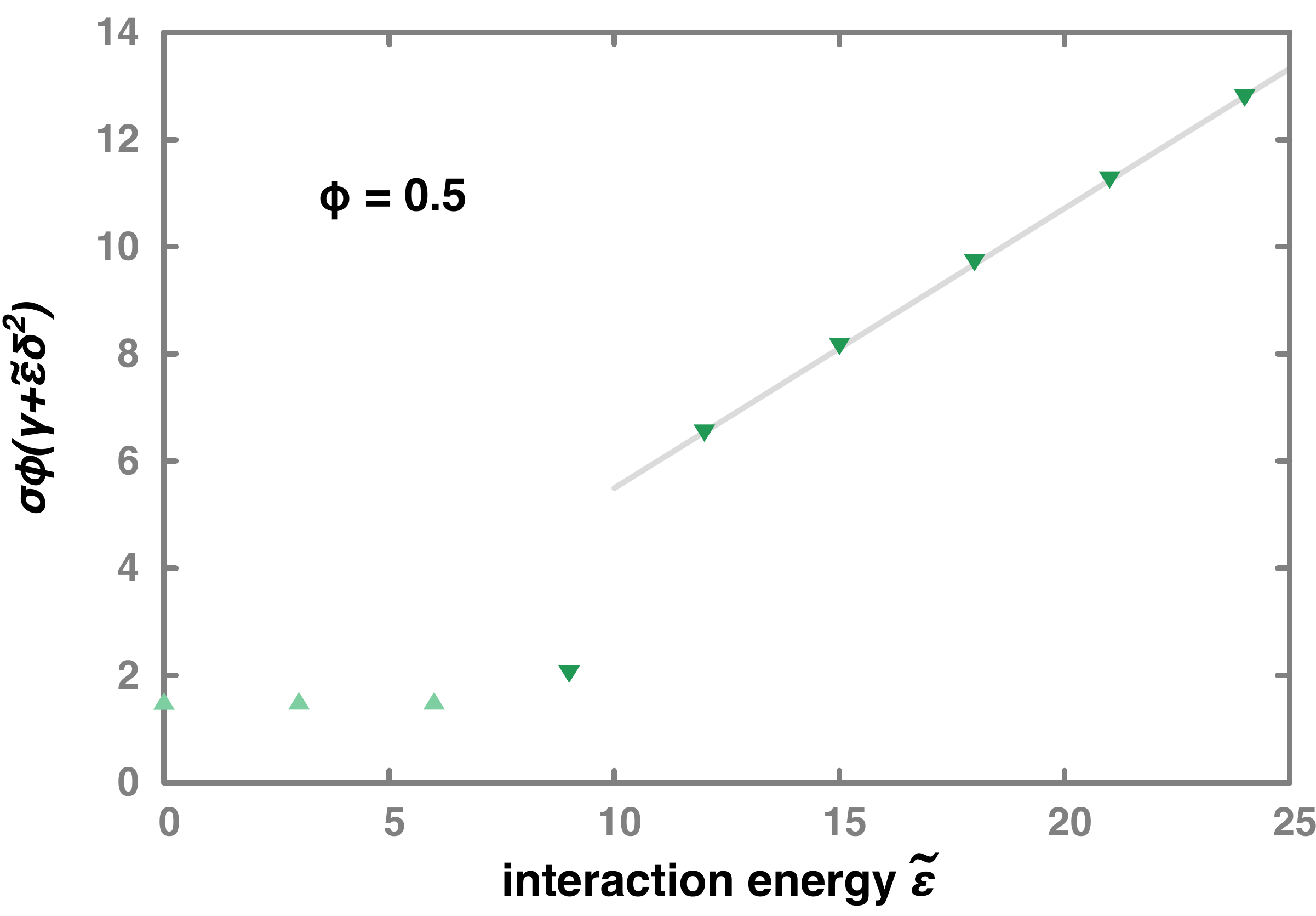}%
 \caption{\label{imgAextra2}Height of the effective smectic ordering potential is given by $\sigma \phi (\gamma + \tilde{\epsilon} \delta ^2)$ as a function of the attraction strength at the volume fraction 0.5. The following phases are identified: smectic A (green triangle up) and smectic A$_2$ (dark green triangle down) phases.The linear dependence of the amplitude of the smectic ordering potential with the attraction strength is also what we find in our simulations. See Figure \ref{imgSmU}(b).}%
 \end{figure}

The height of the effective smectic ordering potential is given by $\sigma \phi (\gamma + \tilde{\epsilon} \delta ^2)$. See Equation \ref{eqP}. This quantity is represented both as a function of the volume fraction for all the attraction strengths investigated (Figure \ref{imgAextra1}) and as a function of the attraction strength for the volume fraction 0.5, in which only the smectic A and smectic A$_2$ phases are found (Figure \ref{imgAextra2}). From the comparison between Figure \ref{imgAextra1} and the inset in Figure \ref{imgSmU} and between Figure \ref{imgAextra2} and the Figure \ref{imgSmU} itself, we find that the theory emulates the linear dependence of the height of the smectic ordering potential as a function of the volume fraction and of the attraction strength except for the angular coefficient depending on the phase. The same is true for the width of the smectic ordering potential, that does not decrease as the attraction strength increases but is constant as can be seen in Figure \ref{imgB}.

\end{appendix}

%

\newpage
\clearpage
\section*{Supplementary Material}
\beginsupplement
\label{sec:sup}
As indicated in the main text, this supplementary material contains (1) the schematic representation of the initial configurations and snapshots of resulting configurations of simulations starting from each of them, (2) the phase diagram for shorter and stiffer chains of $L_0/D=6.46$ and $L_0/L_P=0.1$ that we obtained in preliminary simulations. For the longer and more flexible particles again, we also present (3) further analysis of the average aggregation number, and (4) a comparison between the smectic order parameter for attraction strengths $\epsilon=0\,k_\mathrm{B}T$ (no anti-ferroelectric phase transition in the smectic A phase) and $\epsilon=0.7\,k_\mathrm{B}T$.
 
\begin{figure}[ht]
 \includegraphics[scale=.27]{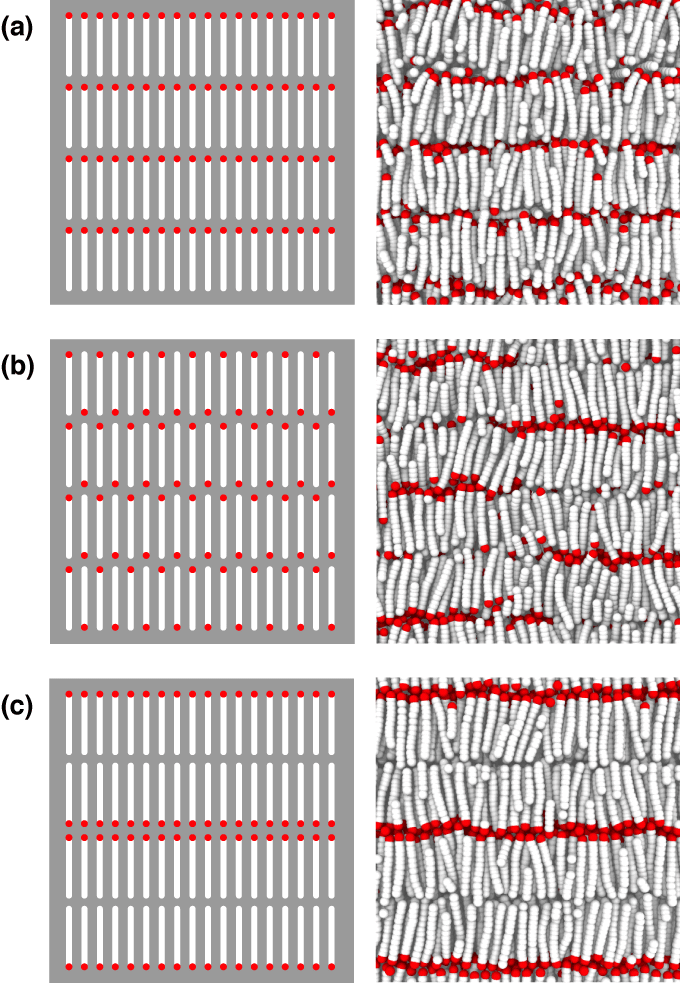}%
 \caption{\label{imgSIsnap}Schematic representation of the initial configurations and snapshots of resulting configurations after 20\,000\,000 timesteps of simulations starting from each of them at approximately constant volume fraction $\phi=0.52$ (at a fixed pressure 1.7\,$\epsilon/D^3$) for the attraction strength of $0.6\,k_\mathrm{B}T$. In the initial configuration in (a), all particles have the same orientations, in (b), particles alternate orientation within the same layer, and in (c), all particles have the same orientation within the layer, but the orientation alternate from consecutive layers.}%
 \end{figure}

\begin{figure}[ht]
 \includegraphics[scale=.34]{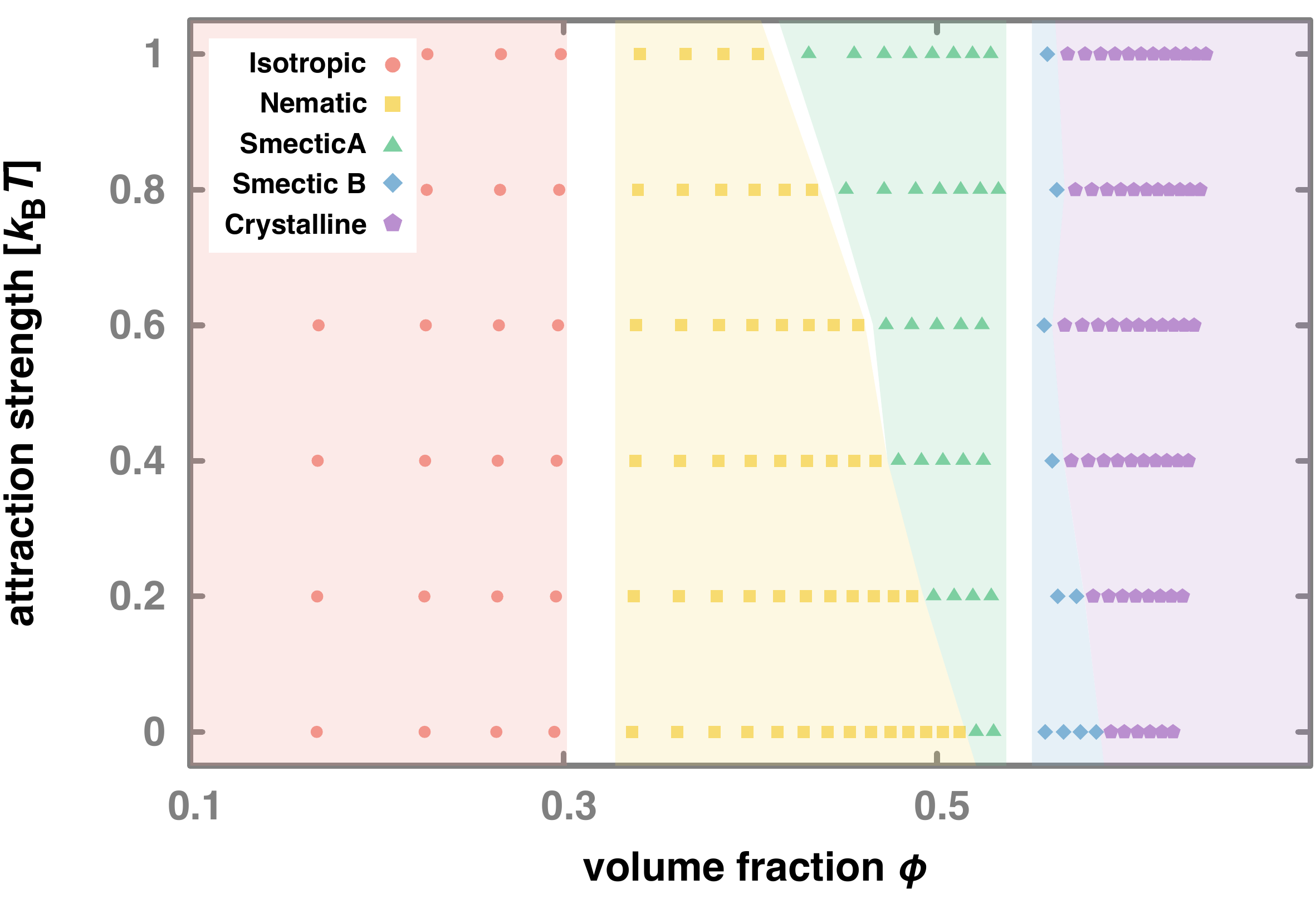}%
 \caption{\label{imgSIpd}Calculated phase diagram of repulsive, rod-like particles that have a single attractive tip as a function of the attraction strength between the end tips (in units of thermal energy) and the volume fraction $\phi$. The particles have a bare aspect ratio $L_0/D = 6.46$ and flexibility of $L_0/L_P = 0.1$. (See main text.) The following phases are identified: isotropic (orange circle), nematic (yellow square), smectic A (green triangle up), smectic B (blue diamond), and crystalline (purple pentagon). We do not check for anti-ferroelectric order.}%
 \end{figure}

 \emph{Initial configurations - }
In preliminary simulations, our particles have an aspect ratio $L_0/D=$ 6.46 and a flexibility $L_0/L_\mathrm{P}=$ 0.1. In Figure \ref{imgSIsnap}, the three different initial configurations (as described in Section II in the main text) are represented. In Figure \ref{imgSIpd}, the schematic representation of the initial configuration is presented followed of a snapshot from the simulations after 20\,000\,000 timesteps at approximately constant volume fraction $\phi=0.52$ (at a fixed pressure 1.7\,$\epsilon/D^3$) for the attraction strength of $0.6\,k_\mathrm{B}T$ for all three cases depicted. 
In (a), all particles have the same orientations in the initial configuration. Note that the usual simulation time of 20\,000\,000 timesteps is insufficient to equilibrate the system. In (b), particles have alternate orientations within the same layer in the initial configuration. In the simulations that have this initial configuration we do observe the formation of the bilayer configuration but they present a wall defect (with shifted bilayer periodicity) that is represented in the snapshot. Finally, in (c), all particles have the same orientation within the layer, but the orientation alternates from consecutive layers.  

\emph{Phase diagram -}
We build the phase diagram for these shorter particles from the third initial configuration as a function of the attraction strength between tips $\epsilon$ and of the volume fraction $\phi$. See Figure \ref{imgSIpd}. In this phase diagram, we do not differentiate between the standard smectic A, smectic B, and crystalline phases and their anti-ferroelectric version, as we do for the phase diagram presented in Figure \ref{imgPD}.
As discussed in the main text, the only impact aspect ratio seems to have is that the volume fractions at which the various phase transitions take place decreases as the aspect ratio increases, as expected from our previous study \cite{Braaf2017}. The same is true for the effect of bending flexibility, which increases the volume fraction at which the various phase transitions occur. 

The next results again apply to particles of aspect ratio $L_0/D=$10.77 and flexibility $L_0/L_\mathrm{P}=$0.3, the same as the ones we describe in the main text.
 
\begin{figure}[ht]
 \includegraphics[scale=.34]{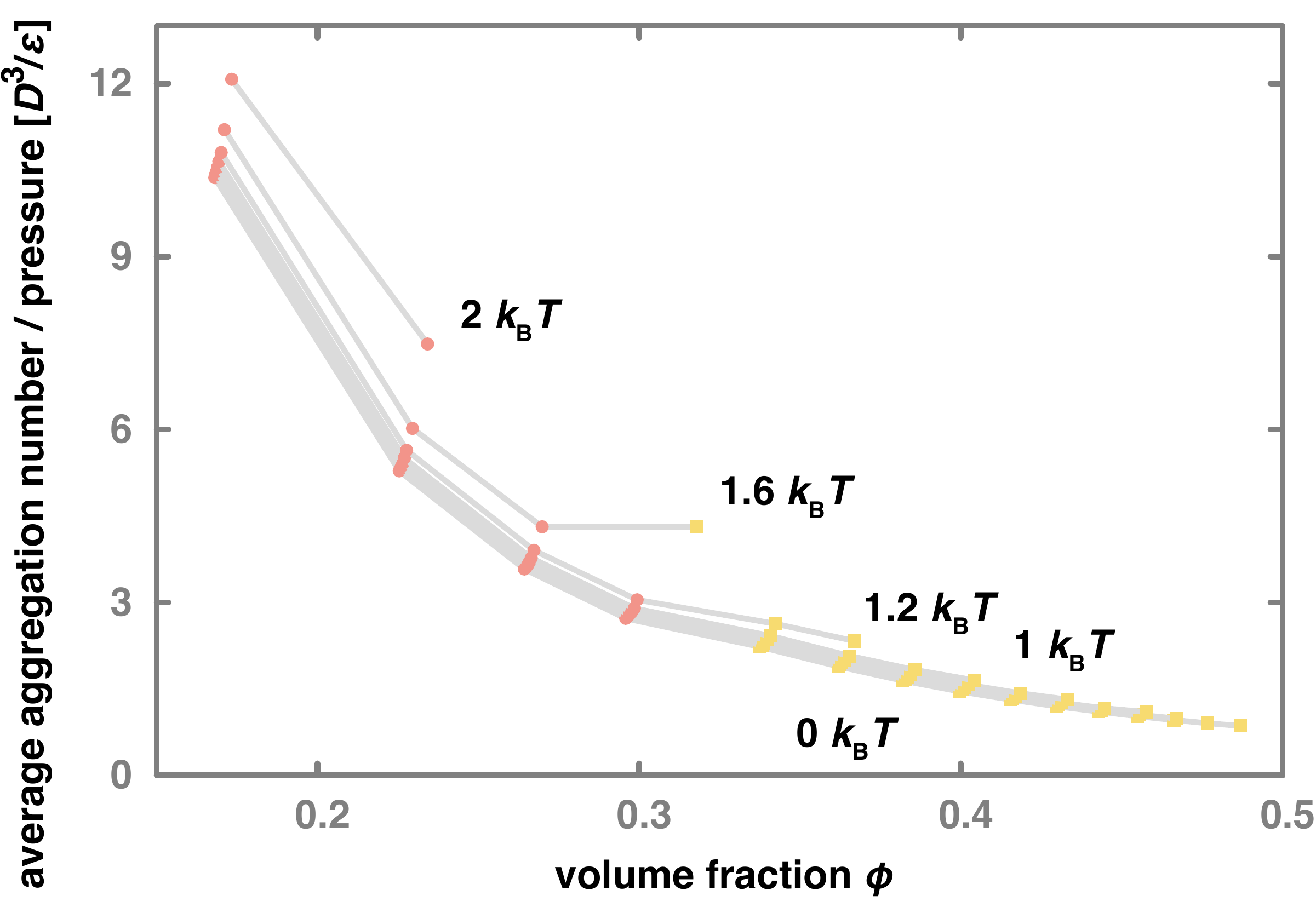}%
 \caption{\label{imgExtra}Average aggregation number for the attraction strengths between 0 and 2\, $k_\mathrm{B}T$ divided by pressure and presented as a function of the volume fraction $\phi$ in the isotropic (orange circle), and nematic (yellow square) phases. The particles have a base aspect ratio $L_0/D = 10.77$ and flexibility of $L/L_P = 0.3$. }%
 \end{figure}

\emph{Cluster statistics - }
Figure \ref{imgExtra} corresponds to another version of the inset of Figure \ref{imgCluster10}(a). 
In this figure, we present the average aggregation number divided by the pressure $P$ in the isotropic and nematic phases for the attraction strengths between 0 and 2\,$k_\mathrm{B}T$, presented as a function of the volume fraction $\phi$. 
The overlapping of the curves for the attraction strengths between 0 and 1\,$k_\mathrm{B}T$ indicates that, at this interval of attraction energies, the aggregation of the attractive tips is mainly due to the concentration.  

  \begin{figure}[ht]
 \includegraphics[scale=.34]{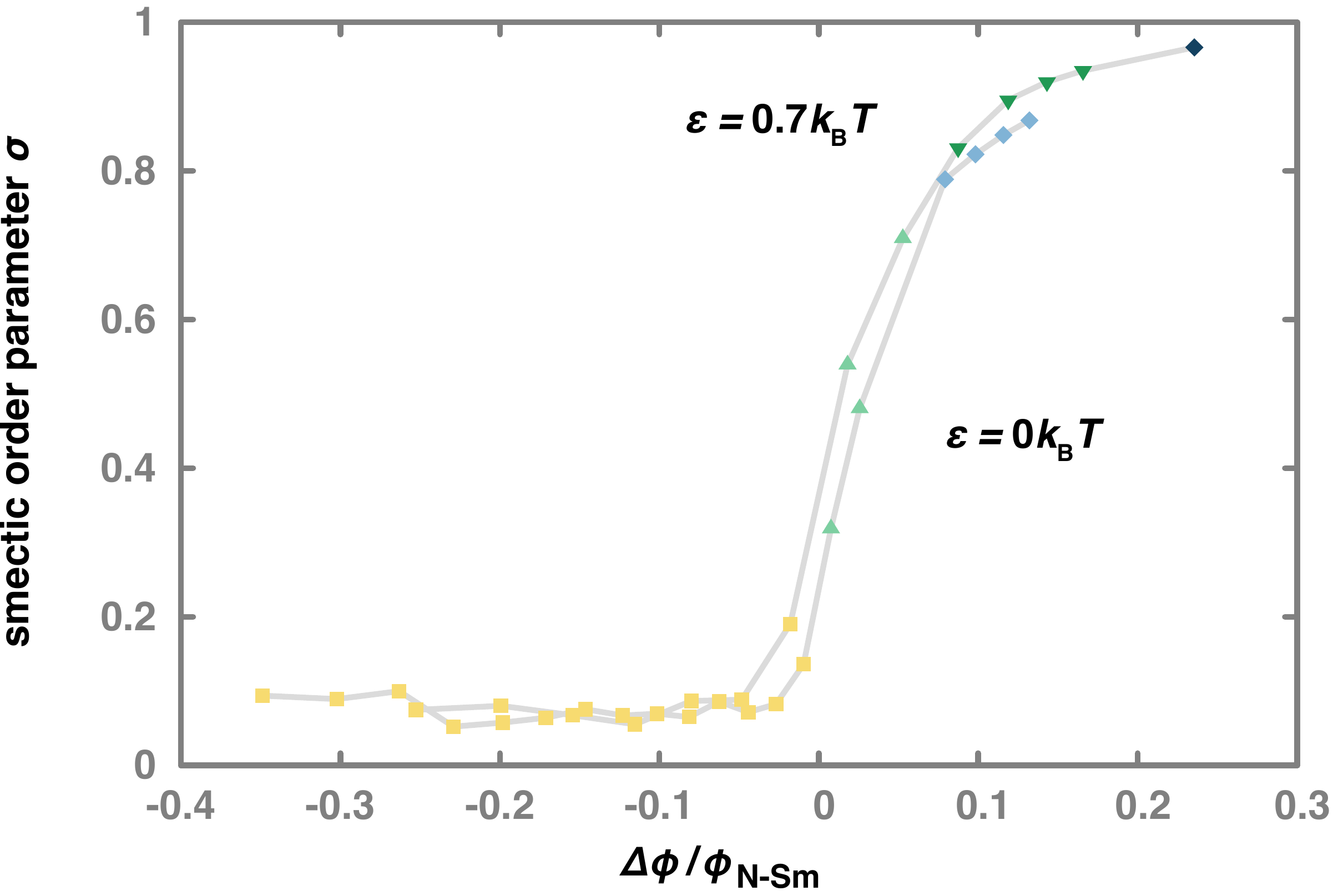}%
 \caption{\label{imgExtra3}Smectic order parameter $\sigma$ of repulsive, rod-like particles that have a single attractive tip as a function of $\Delta \phi / \phi_{N-Sm}$, where $\Delta \phi$ is the difference between the volume fraction and the volume fraction at the Nematic-to-Smectic A (for $\epsilon=0$) or Smectic A$_2$ phase (for $\epsilon=0.7\,k_\mathrm{B}T$) transition,  $\Delta \phi = \phi - \phi_{N-Sm}$, for attraction strengths of 0 and $0.7\,k_\mathrm{B}T$. Particles have aspect ratio $L_0/D = 10.77$ and flexibility of $L_0/L_P = 0.3$. The following phases are identified: nematic (yellow square), smectic A (green triangle up), smectic A$_2$ (dark green triangle down), smectic B  (blue diamond), and smectic B$_2$ (dark blue diamond) phases.}%
 \end{figure}
 
\emph{Anti-ferroelectric phase transition - } 
Figure \ref{imgExtra3} is another version of the inset of Figure \ref{imgAnti}. There, we compare the smectic order parameter $\sigma$ as a function of $\Delta \phi / \phi_{N-Sm}$, where $\Delta \phi$ is the difference between the volume fraction and the volume fraction at the Nematic-to-Smectic A (for $\epsilon=0\,k_\mathrm{B}T$) or Smectic A$_2$ phase (for $\epsilon=0.7\,k_\mathrm{B}T$) transition,  $\Delta \phi = \phi - \phi_{N-Sm}$. The volume fraction $\phi_{N-Sm}$ at the transition is estimated by the average between the highest and lowest values of the volume fraction at which, respectively, the nematic and smectic A or smectic A$_2$ phases are stable. We find that the smectic order parameter $\sigma$ is smaller at smaller attraction strength from the comparison of the values of $\epsilon$ between 0 and 0.7$\,k_\mathrm{B}T$.
Therefore, this result shows that the degree of order increases due to the attraction strength between single tips in the layered phases.

\end{document}